\documentclass[journal]{IEEEtran}
\usepackage[OT1]{fontenc}
\usepackage[utf8]{inputenc}
\usepackage{bm}
\usepackage{amsmath}
\usepackage{amssymb}
\usepackage{graphicx}
\makeatletter
\IEEEoverridecommandlockouts

\usepackage{color}
\usepackage{bm}
\usepackage{algorithm}
\usepackage{algorithmic}
\usepackage{multirow}
\usepackage{booktabs}
\usepackage{array}
\usepackage{amsthm}
\usepackage{lipsum}
\usepackage{enumerate}
\usepackage{stfloats}
\usepackage{subfigure}
\usepackage{cases}
\usepackage{diagbox}
\usepackage{threeparttable}
\usepackage{cite}
\usepackage{url}

\newtheorem{assump}{Assumption}\newtheorem{remark}{Remark}

\title{Active beyond-diagonal Reconfigurable Intelligent Surfaces: Modeling, Architecture Design,\\ and Optimization}

\author{Shanpu Shen, \textit{Senior Member, IEEE}, Hongyu Li, \textit{Member, IEEE},\\ Matteo Nerini, \textit{Member, IEEE}, and Bruno Clerckx, \textit{Fellow, IEEE}
\thanks{S. Shen is with the Department of Electrical Engineering and Electronics, University of Liverpool, Liverpool L69 3GJ, U.K. (e-mail: Shanpu.Shen@liverpool.ac.uk).}
\thanks{H. Li is with the Internet of Things Thrust, The Hong Kong University of Science and Technology (Guangzhou), Guangzhou 511400, China (e-mail: hongyuli@hkust-gz.edu.cn)}
\thanks{M. Nerini and B. Clerckx are with the Department of Electrical and Electronic Engineering, Imperial College London, London SW7 2AZ, U.K. (e-mail:\{m.nerini20,b.clerckx\}@imperial.ac.uk).}
}

\makeatother

\begin{document}
\title{Active Beyond-Diagonal Reconfigurable Intelligent Surfaces: Modeling,
Architecture Design, \\
 and Optimization}
\author{Shanpu Shen,~\IEEEmembership{Senior Member,~IEEE}, Hongyu Li, \IEEEmembership{Member,~IEEE},
\\
 Matteo Nerini, \IEEEmembership{Member,~IEEE}, Qingqing Wu,~\IEEEmembership{Senior Member,~IEEE},
Bruno Clerckx,~\IEEEmembership{Fellow,~IEEE}\thanks{Manuscript received; 
The work of S. Shen was funded by the National Natural Science Foundation of China under Grant 62601016,	the Science and Technology
Development Fund, Macau SAR (File/Project no. 001/2024/SKL and 0002/2025/EQP), by University of Macau (File no. SRG2025-00060-IOTSC), and by 
University of Macau Development Foundation (UMDF) (File no. UMDF-TISF-I/2026/025/IOTSC). The work of H. Li was funded by the National Natural Science Foundation of China under Grant 62501509 and the Natural Science Foundation of Guangdong Province under Grant 2026A1515011048.
The work of Q. Wu was funded by the National Natural Science Foundation of China under Grant 62371289. \textit{(Corresponding
author: Hongyu Li.)}}\thanks{S. Shen is with the State Key Laboratory of Internet of Things for
Smart City and Department of Electronic and Communication Engineering,
University of Macau, Macau, China (e-mail: shanpushen@um.edu.mo).}\thanks{H. Li is with the Internet of Things Thrust, The Hong Kong University
of Science and Technology (Guangzhou), Guangzhou 511400, China (email:
hongyuli@hkust-gz.edu.cn).}\thanks{Q. Wu is with the Department of Electronic Engineering, Shanghai Jiao
Tong University, Shanghai 200240, China (email: qingqingwu@sjtu.edu.cn).}\thanks{M. Nerini, and B. Clerckx are with the Department of Electrical and
Electronic Engineering, Imperial College London, London SW7 2AZ, U.K.
(e-mail: \{m.nerini20, b.clerckx\}@imperial.ac.uk).}}
\maketitle
\begin{abstract}
Beyond-diagonal reconfigurable intelligent surfaces (BD-RISs) are
an emerging RIS 2.0 technology for future wireless communication.
However, BD-RISs are primarily passive without active amplification,
suffering from severe multiplicative path loss. To address the concern
of multiplicative path loss, in this work we investigate the active
BD-RIS including the modeling, architecture design, and optimization.
We first analyze the active BD-RIS using multiport network theory
with scattering parameters and derive a physical and electromagnetic
compliant active BD-RIS aided communication model. We also design
two new active BD-RIS architectures, namely fully- and group-connected
active BD-RISs. Based on the proposed model and architecture, we investigate
the active BD-RIS aided single-input single-output system and derive
the closed-form optimal solution and scaling law of the signal-to-noise
ratio. We further investigate the active BD-RIS aided multiple-input
multiple-output system and propose an iterative algorithm based on
quadratically constrained quadratic programming to maximize the spectral
efficiency. Numerical results are provided and show that the active
BD-RIS can achieve higher spectral efficiency than the active/passive
diagonal RIS and passive BD-RIS. For example, to achieve the same
spectral efficiency, the number of elements required by active BD-RIS
is less than half of that required by active diagonal RIS, showing
the advantages of active BD-RIS.
\end{abstract}


\section{Introduction}

\label{sec:intro}

\IEEEPARstart{R}{econfigurable} intelligent surfaces (RISs) have
gained significant attentions in the past years as one of the key
technologies to enable the promise of the sixth generation (6G) communication
\cite{Mu2024}. An RIS is made up of a large number of reconfigurable
elements, so that it can smartly reconfigure the wireless propagation
environment by collaboratively controlling the RIS elements \cite{di2020smart}.
In the conventional RIS architecture, each RIS element is connected
to a reconfigurable impedance component without interconnection to
other elements and it is characterized by a diagonal scattering matrix,
which is therefore referred to as single-connected RIS or diagonal
RIS (D-RIS). However, D-RISs (in the lossless form) can only control
the phase of the scattered wave and provides limited degrees of freedom
for wave manipulation.

To break this limitation, a more general RIS 2.0, called beyond-diagonal
RIS (BD-RIS), has been developed by interconnecting RIS elements through
variable impedance \cite{li2023reconfigurable}. A BD-RIS is characterized
by a scattering matrix, not limited to being diagonal, which allows
controlling not only the phase but also the magnitude of the scattered
wave and thus providing more degrees of freedom for enhancing wireless
systems \cite{Li2025c}. Depending on how the RIS elements are interconnected,
there are different BD-RIS architectures, such as fully-connected
RIS \cite{shen2021}, where the RIS elements are all interconnected,
and group-connected RIS \cite{shen2021}, \cite{Li2023a}, \cite{Nerini2024c},
where only the RIS elements in the same group are interconnected.
To achieve a better tradeoff between circuit complexity and performance,
the tree-, forest-, band-, and stem-connected, characterized by graph
theory, have been proposed and developed in \cite{Nerini2024b}, \cite{Wu2025a}.
Benefiting from the interconnection between RIS elements, BD-RISs
with hybrid transmitting and reflecting mode have been developed and
proposed, which supports the simultaneous transmission and reflection
of signal to broaden the coverage \cite{li2022beyond}. Moreover,
extending the hybrid transmitting and reflecting mode, a more general
BD-RIS with multi-sector mode has been proposed, where the signal
impinging into one sector can be re-radiated by the other sector,
enabling not only the full-space coverage but also high channel gain
\cite{Li2023}. With these benefits, BD-RIS has been adopted to improve
different wireless systems and scenarios including wireless power
transfer \cite{Azarbahram2025a}, integrated sensing and communication
(ISAC) \cite{Wang2024e}, \cite{Liu2024a}, \cite{Guang2024}, rate-splitting
multiple access (RSMA) \cite{Soleymani2023}, \cite{Li2024c}, multi-cell
multiple-input multiple-output (MIMO) \cite{DeSena2024}, cell-free
massive MIMO \cite{Hua2024}, \cite{Li2025b}, orthogonal frequency
division multiplexing (OFDM) \cite{Li2025}, etc. In addition, new
channel estimation methods tailored for BD-RIS have been developed
\cite{Li2024a}, \cite{Wang2025} and a hybrid transmitting and reflecting
BD-RIS with independent beam control and power splitting has been
designed, prototyped, and experimented \cite{Ming2025a}. All these
works have shown the superior performance of BD-RIS compared with
conventional D-RISs. Nevertheless, BD-RISs in the previous literature
are primarily passive without any active amplification, which is similar
to passive D-RIS and thus suffers from severe multiplicative path
loss for the cascaded channel. Consequently, the performance gain
of passive BD-RIS heavily relies on an extremely large number of elements
to compensate the multiplicative path loss, which increases the size,
complexity, and cost and therefore poses a challenge for practical
utility.

To address the concern of multiplicative path loss, active RISs have
been proposed and developed \cite{Long2021}, \cite{Khoshafa2021}.
In contrast with passive RISs which can only adjust the phase of scattered
wave by passive variable impedance, active RISs can not only adjust
the phase but also actively amplify the amplitude with a gain larger
than one \cite{Ahmed2025}, by integrating reflection-type amplifiers
within each RIS element \cite{Zhang2023} or a subset of RIS elements
\cite{Liu2022b}. By simultaneously adjusting the phase and amplifying
the magnitude, active RISs are able to reconfigure the propagation
environment while compensating the multiplicative path loss, so as
to effectively enhance the performance of wireless systems such as
spectral efficiency \cite{Liu2022c} and energy efficiency \cite{Ma2023}.
Leveraging such capability, active RISs have been applied to improve
different wireless systems and scenarios such as simultaneous wireless
information and power transfer \cite{Zargari2022}, \cite{Ren2023},
non-orthogonal multiple access \cite{Chen2024a}, \cite{Gong2024},
mobile edge computing \cite{Peng2022}, \cite{Li2024d}, RSMA \cite{Niu2023},
OFDM \cite{Chian2024}, \cite{Zhang2023a}, ISAC \cite{Salem2023},
\cite{Sun2024a}, and UAV \cite{Peng2024a}. Although these works
have demonstrated the advantages of active RIS, there is no interconnection
between RIS elements in the considered active RIS architecture, so
that they all fall into the category of active D-RIS, which is characterized
by a diagonal matrix and has limited degrees of freedom. Therefore,
it is necessary to explore the active BD-RIS architecture with element
interconnection to provide more degrees of freedom to enhance the
wireless systems.

In this work, we investigate the active BD-RIS including the modeling,
architecture design, and optimization. Compared with the active D-RIS,
the proposed active BD-RIS introduces controllable interconnection
between RIS elements, so that it is characterized by a beyond-diagonal
scattering matrix and has more degrees of freedom to enhance wireless
systems. The contributions of this work are summarized as follows.

\textit{First}, we analyze the active BD-RIS aided wireless communication
systems using multiport network theory with scattering parameters
and then derive a physical and electromagnetic (EM) compliant active
BD-RIS aided communication model. Utilizing the multiport network
theory with scattering parameters can accurately model the scattering
mechanism of active BD-RIS, particularly the active amplification
with dynamic noise. The derived communication model is general to
take into account the active reconfigurable impedance network of active
BD-RIS, impedance mismatching, and mutual coupling, while the conventional
active RIS model is a particular instance of the proposed active BD-RIS
model. In addition, we analyze the relationship between active and
passive BD-RISs and show that the derived active BD-RIS model becomes
the passive BD-RIS model when replacing the reflection-type amplifiers
with fixed reactive loads.

\textit{Second}, we design two new active BD-RIS architectures, called
fully-connected active BD-RIS and group-connected active BD-RIS, which
are respectively characterized by a complex symmetric matrix and a
block diagonal matrix with each block being complex symmetric. In
contrast to the conventional active D-RIS, which is referred to as
single-connected active BD-RIS, the proposed fully- and group-connected
active BD-RIS allows the signal to flow through different elements
by the controllable interconnection, which is more general and provides
more degrees-of-freedom to manipulate the wave and improve the wireless
systems.

\textit{Third}, we investigate the active BD-RIS aided single-input
single-output (SISO) system and derive the closed-form optimal solution
and scaling law of the signal-to-noise ratio (SNR) in comparison with
active D-RIS and passive BD-RIS. Specifically, considering a simplified
case with equal amplification factor and no direct channel, we derive
the closed-form optimal solution for the fully- and group-connected
active BD-RIS and the corresponding SNR. Based on this, we derive
the scaling law of the SNR as a function of the number of elements,
comparing with the active D-RIS and passive BD-RIS. The scaling law
shows that, with equal amplification factor, the maximum SNR gain
of active BD-RIS over active D-RIS in SISO system can be up to 1.62,
which is same as the maximum SNR gain of passive BD-RIS over passive
D-RIS. It also shows that the active BD-RIS maintains its benefit
over passive BD-RIS until an extremely large number of elements.

\textit{Fourth}, we investigate the active BD-RIS aided MIMO systems
and propose an iterative algorithm to maximize the spectral efficiency.
The proposed algorithm is based on iteratively optimizing the precoder
and the scattering matrix of active BD-RIS by quadratically constrained
quadratic programming (QCQP). The convergence and computational complexity
are both analyzed.

\textit{Fifth}, we evaluate the performance of the active BD-RIS aided
MIMO systems, compared with active/passive D-RIS and passive BD-RIS.
The numerical results show that the active BD-RIS can achieve higher
spectral efficiency than the active/passive D-RIS and passive BD-RIS.
For example, to achieve the same spectral efficiency, the number of
elements required by active BD-RIS is less than half of that required
by active D-RIS, showing the advantages of active BD-RIS.

\textit{Organization}: Section II presents the multiport network analysis
with scattering parameters and develops the fully- and group-connected
active BD-RISs. Section III provides the active BD-RIS aided communication
model. Section IV investigates the active BD-RIS aided SISO system.
Section V investigates the active BD-RIS aided MIMO system. Section
VI evaluates the performance of the active BD-RIS and Section VII
provides conclusions of this work.

\textit{Notations:} Boldface lower- and upper-case letters indicate
column vectors and matrices, respectively. $(\cdot)^{\mathsf{T}}$,
$(\cdot)^{*}$, $(\cdot)^{\mathsf{H}}$, and $(\cdot)^{-1}$ denote
the transpose, conjugate, conjugate-transpose, and inverse operations,
respectively. $\mathbb{C}$ and $\mathbb{R}$ denote the sets of complex
and real numbers, respectively. $\mathbb{E}$ denotes the statistical
expectation. $\Re\{\cdot\}$ denotes the real part of complex numbers.
$\mathsf{blkdiag}(\cdot)$ represents a block-diagonal matrix and
$\mathsf{diag}(\cdot)$ represents a diagonal matrix. $\|\cdot\|_{2}$
and $\|\cdot\|_{\mathsf{F}}$ denote the $\ell_{2}$ norm and Frobenius
norm, respectively. $\jmath=\sqrt{-1}$ denotes the imaginary unit.
$\mathsf{Tr}(\cdot)$ denotes the trace of a matrix. $\mathbf{I}_{N}$
denotes an $N\times N$ identity matrix. $\mathbf{0}$ denotes an
all-zero matrix. $a\sim\mathcal{CN}(0,\sigma^{2})$ characterizes
the circular symmetric complex Gaussian distribution. $[\mathbf{A}]_{i:i',j:j'}$
extracts the $i$th to $i'$th rows and the $j$th to $j'$th columns
of $\mathbf{A}$.

\section{Multiport Network Analysis}

\label{sec:syst_mod}

As the starting point, we adopt the multiport network theory to analyze
and model the active BD-RIS aided wireless communication systems in
this section.

\begin{figure}
\centering \includegraphics[width=0.48\textwidth]{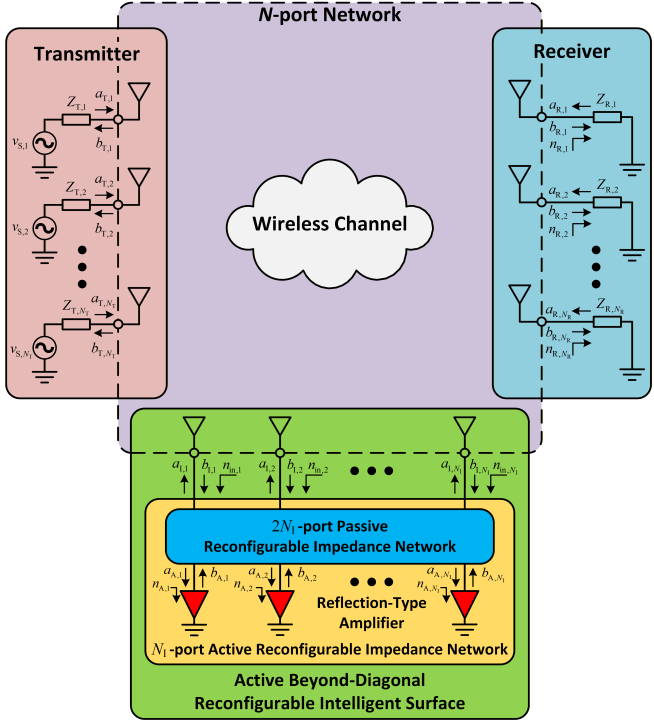}
\caption{Diagram of active BD-RIS aided wireless communication system.}
\label{fig:model}
\end{figure}

Consider an active BD-RIS aided multi-antenna system as illustrated
in Fig. \ref{fig:model}, consisting of an $N_{\mathrm{T}}$-antenna
transmitter, an $N_{\mathrm{R}}$-antenna receiver, and an $N_{\mathrm{I}}$-element
active BD-RIS. The whole system has $N=N_{\mathrm{T}}+N_{\mathrm{I}}+N_{\mathrm{R}}$
antennas, so that we can model it as an $N$-port network. Using the
scattering parameters, we can characterize the $N$-port network by
a scattering matrix $\mathbf{S}\in\mathbb{C}^{N\times N}$, relating
the incident wave $\mathbf{a}\in\mathbb{C}^{N\times1}$ and reflected
wave $\mathbf{b}\in\mathbb{C}^{N\times1}$ at the $N$ ports as $\mathbf{b}=\mathbf{S}\mathbf{a}$.
Introducing the subscripts T, I, and R to respectively denote the
transmitter, active BD-RIS, and receiver, we can partition $\mathbf{a}$,
$\mathbf{b}$, and $\mathbf{S}$ as 
\begin{equation}
\mathbf{a}=\left[\begin{array}{l}
\mathbf{a}_{\mathrm{T}}\\
\mathbf{a}_{\mathrm{I}}\\
\mathbf{a}_{\mathrm{R}}
\end{array}\right],\:\mathbf{b}=\left[\begin{array}{l}
\mathbf{b}_{\mathrm{T}}\\
\mathbf{b}_{\mathrm{I}}\\
\mathbf{b}_{\mathrm{R}}
\end{array}\right],\mathbf{S}=\left[\begin{array}{ccc}
\mathbf{S}_{\mathrm{TT}} & \mathbf{S}_{\mathrm{TI}} & \mathbf{S}_{\mathrm{TR}}\\
\mathbf{S}_{\mathrm{IT}} & \mathbf{S}_{\mathrm{II}} & \mathbf{S}_{\mathrm{IR}}\\
\mathbf{S}_{\mathrm{RT}} & \mathbf{S}_{\mathrm{RI}} & \mathbf{S}_{\mathrm{RR}}
\end{array}\right],
\end{equation}
where $\mathbf{a}_{i}=[a_{i,1},\ldots,a_{i,N_{i}}]^{\mathsf{T}}\in\mathbb{C}^{N_{i}\times1}$
and $\mathbf{b}_{i}=[b_{i,1},\ldots,b_{i,N_{i}}]^{\mathsf{T}}\in\mathbb{C}^{N_{i}\times1}$
for $i\in\left\{ \mathrm{T},\mathrm{I},\mathrm{R}\right\} $ are the
incident and reflected waves of the antennas at transmitter/active
BD-RIS/receiver, respectively. $\mathbf{S}_{ii}\in\mathbb{C}^{N_{i}\times N_{i}}$
is the scattering matrix of the antenna array at transmitter/active
BD-RIS/receiver. $\mathbf{S}_{ij}\in\mathbb{C}^{N_{i}\times N_{j}}$
is the transmission scattering matrices, $\forall i\ne j,i,j\in\left\{ \mathrm{T},\mathrm{I},\mathrm{R}\right\} $,
e.g. from transmitter to receiver.

\subsection{Transmitter and Receiver}

At the transmitter, the $i$th transmit antenna for $i=1,\ldots,N_{\mathrm{T}}$
is in series with a voltage source $v_{\mathrm{S},i}$, and a source
impedance $Z_{\mathrm{T},i}$. Thus, we can relate $\mathbf{a}_{\mathrm{T}}$
and $\mathbf{b}_{\mathrm{T}}$ by 
\begin{align}
\mathbf{a}_{\mathrm{T}} & =\mathbf{b}_{\mathrm{S},\mathrm{T}}+\mathbf{\Gamma}_{\mathrm{T}}\mathbf{b}_{\mathrm{T}},\label{eq:tx}
\end{align}
where $\mathbf{b}_{\mathrm{S},\mathrm{T}}$ refers to the wave source
vector and $\mathbf{\Gamma}_{\mathrm{T}}\in\mathbb{C}^{N_{\mathrm{T}}\times N_{\mathrm{T}}}$
refers to the scattering matrix for the $N_{\mathrm{T}}$ source impedance.
Denoting $\mathbf{v}_{\mathrm{S},\mathrm{T}}=[v_{\mathrm{S},1},\ldots,v_{\mathrm{S},N_{\mathrm{T}}}]^{\mathsf{T}}\in\mathbb{C}^{N_{\mathrm{T}}\times1}$
as the source voltage vector, we have that $\mathbf{b}_{\mathrm{S},\mathrm{T}}=\frac{\mathbf{I}_{N_{\mathrm{T}}}-\mathbf{\Gamma}_{\mathrm{T}}}{2}\mathbf{v}_{\mathrm{S},\mathrm{T}}$.
$\mathbf{\Gamma}_{\mathrm{T}}$ is diagonal and its $\left(i,i\right)$th
entry is the reflection coefficient for the $i$th source impedance,
that is $\left[\mathbf{\Gamma}_{\mathrm{T}}\right]_{i,i}=\frac{Z_{\mathrm{T},i}-Z_{0}}{Z_{\mathrm{T},i}+Z_{0}}$,
where $Z_{0}$ is the reference impedance, e.g. $Z_{0}=50\:\Omega$.

At the receiver, the $i$th receive antenna, for $i=1,\ldots,N_{\mathrm{R}}$,
is in series with a load impedance $Z_{\mathrm{R},i}$. Thus, we can
relate $\mathbf{a}_{\mathrm{R}}$ and $\mathbf{b}_{\mathrm{R}}$ by
\begin{align}
\mathbf{a}_{\mathrm{R}} & =\mathbf{\Gamma}_{\mathrm{R}}\left(\mathbf{b}_{\mathrm{R}}+\mathbf{n}_{\mathrm{R}}\right),\label{eq:rx}
\end{align}
where $\mathbf{n}_{\mathrm{R}}\sim\mathcal{CN}(\mathbf{0},\sigma_{\mathrm{R}}^{2}\mathbf{I}_{N_{\mathrm{R}}})$
refers to the noise at the receiver having power $\sigma_{\mathrm{R}}^{2}$
and $\mathbf{\Gamma}_{\mathrm{R}}\in\mathbb{C}^{N_{\mathrm{R}}\times N_{\mathrm{R}}}$
refers to the scattering matrix for the $N_{\mathrm{R}}$ load impedance.
$\mathbf{\Gamma}_{\mathrm{R}}$ is a diagonal matrix and its $\left(i,i\right)$th
entry is the reflection coefficient for the $i$th load impedance,
that is $\left[\mathbf{\Gamma}_{\mathrm{R}}\right]_{i,i}=\frac{Z_{\mathrm{R},i}-Z_{0}}{Z_{\mathrm{R},i}+Z_{0}}$.

\subsection{Active Beyond-Diagonal Reconfigurable Intelligent Surface}

An $N_{\mathrm{I}}$-element active BD-RIS consists of $N_{\mathrm{I}}$
antennas, a $2N_{\mathrm{I}}$-port passive reconfigurable impedance
network, and $N_{\mathrm{I}}$ reflection-type amplifiers. Specifically,
the $N_{\mathrm{I}}$ antennas are connected to ports 1-$N_{\mathrm{I}}$
of the $2N_{\mathrm{I}}$-port passive reconfigurable impedance network,
where the remaining ports $(N_{\mathrm{I}}+1)$-$2N_{\mathrm{I}}$
are connected to the $N_{\mathrm{I}}$ reflection-type amplifiers,
as illustrated in Fig. \ref{fig:model}. Therefore, the $N_{\mathrm{I}}$-element
active BD-RIS can be also viewed as $N_{\mathrm{I}}$ antennas connected
to an $N_{\mathrm{I}}$-port active reconfigurable impedance network,
which is made up of the $2N_{\mathrm{I}}$-port passive reconfigurable
impedance network and the $N_{\mathrm{I}}$ reflection-type amplifiers. In this model, the reflection-type amplifiers are the only active part of the active BD-RIS and the rest is entirely passive.  

The $2N_{\mathrm{I}}$-port passive reconfigurable impedance network
can be characterized by its scattering matrix $\mathbf{\Phi}\in\mathbb{C}^{2N_{\mathrm{I}}\times2N_{\mathrm{I}}}$.
Accordingly, we can partition $\mathbf{\Phi}$ as 
\begin{equation}
\mathbf{\Phi}=\left[\begin{matrix}\mathbf{\Phi}_{\mathrm{II}} & \mathbf{\Phi}_{\mathrm{IA}}\\
\mathbf{\Phi}_{\mathrm{AI}} & \mathbf{\Phi}_{\mathrm{AA}}
\end{matrix}\right],
\end{equation}
where $\mathbf{\Phi}_{\mathrm{II}}\in\mathbb{C}^{N_{\mathrm{I}}\times N_{\mathrm{I}}}$
is the scattering matrix for ports 1-$N_{\mathrm{I}}$ that are connected
to antennas, $\mathbf{\Phi}_{\mathrm{AA}}\in\mathbb{C}^{N_{\mathrm{I}}\times N_{\mathrm{I}}}$
is the scattering matrix for ports $(N_{\mathrm{I}}+1)$-$2N_{\mathrm{I}}$
that are connected to amplifiers, and $\mathbf{\Phi}_{\mathrm{AI}},\mathbf{\Phi}_{\mathrm{IA}}\in\mathbb{C}^{N_{\mathrm{I}}\times N_{\mathrm{I}}}$
is the transmission scattering matrices from ports 1-$N_{\mathrm{I}}$
to ports $(N_{\mathrm{I}}+1)$-$2N_{\mathrm{I}}$ and vice versa,
respectively. Therefore, we have 
\begin{equation}
\left[\begin{matrix}\mathbf{a}_{\mathrm{I}}\\
\mathbf{a}_{\mathrm{A}}
\end{matrix}\right]=\left[\begin{matrix}\mathbf{\Phi}_{\mathrm{II}} & \mathbf{\Phi}_{\mathrm{IA}}\\
\mathbf{\Phi}_{\mathrm{AI}} & \mathbf{\Phi}_{\mathrm{AA}}
\end{matrix}\right]\left[\begin{matrix}\mathbf{b}_{\mathrm{I}}+\mathbf{n}_{\mathrm{in}}\\
\mathbf{b}_{\mathrm{A}}
\end{matrix}\right],\label{eq:a-b_RIS}
\end{equation}
where $\mathbf{n}_{\mathrm{in}}\in\mathbb{C}^{N_{\mathrm{I}}\times1}$
is the input noise with power $\sigma_{\mathrm{in}}^{2}=kT_{\mathrm{in}}B$
where $k$ is the Boltzmann's constant, $T_{\mathrm{in}}$ is the
noise temperature of the input noise, and $B$ is the bandwidth, so
that we have $\mathbf{n}_{\mathrm{in}}\sim\mathcal{CN}(\mathbf{0},\sigma_{\mathrm{in}}^{2}\mathbf{I}_{N_{\mathrm{I}}})$.
$\mathbf{a}_{\mathrm{A}}=[a_{\mathrm{A},1},\ldots,a_{\mathrm{A},N_{\mathrm{I}}}]^{\mathsf{T}}\in\mathbb{C}^{N_{\mathrm{I}}\times1}$
and $\mathbf{b}_{\mathrm{A}}=[b_{\mathrm{A},1},\ldots,b_{\mathrm{A},N_{\mathrm{I}}}]^{\mathsf{T}}\in\mathbb{C}^{N_{\mathrm{I}}\times1}$
respectively refer to the incident waves to and reflected waves from
the $N_{\mathrm{I}}$ amplifiers, and are further related to each
other by 
\begin{equation}
\mathbf{b}_{\mathrm{A}}=\mathbf{A}\mathbf{a}_{\mathrm{A}}+\mathbf{A}\mathbf{n}_{\mathrm{A}},\label{eq:b-a_PA}
\end{equation}
where $\mathbf{A}=\mathsf{diag}(A_{1},\ldots,A_{N_{\mathrm{I}}})\in\mathbb{R}^{N_{\mathrm{I}}\times N_{\mathrm{I}}}$
with $A_{i}\ge 0$ denoting the amplification factor for
the $i$th amplifier\footnote{In this work we consider an ideal case that all amplifiers work in their linear region and have no amplification limitations. This is generally true since the incident wave to amplifiers are weak due to the path loss of transmitter-RIS channel. To address the concern of non-linearity of practical amplifiers, we can either bound the amplification factor, which also guarantees stability as \cite{Gavriilidis2025}, or introduce non-linear models, where $\mathbf{A}$ will be a function of $\mathbf{a}_\mathrm{I}$ \cite{yumeng2023}. In this sense a new model explicitly capturing the bounded $\mathbf{A}$ or the relationship between $\mathbf{A}$ and $\mathbf{a}_\mathrm{I}$ needs to be derived.}. Specifically, the amplification factor\footnote{The phase of the reflection coefficient $A_i$ can be combined with the passive reconfigurable impedance network such that we can simply consider the impact of its amplitude.} of the reflection-type amplifier is essentially its reflection coefficient given by $A_i = \frac{Z_{\mathrm{A},i} - Z_0}{Z_{\mathrm{A},i} + Z_0}$, where $Z_{\mathrm{A},i}\in\mathbb{C}$ is its input impedance. When $\Re\{Z_{\mathrm{A},i}\} <0$, we have $|A_i|>1$, meaning that the reflected wave is amplified relative to the incident wave. This can be implemented using tunnel diode \cite{francesco2015,jongwon2017,Seiran2017}, transistor \cite{rao2023}, or a varactor diode and a coupling circuit with a pump input \cite{Zhang2023}.
$\mathbf{n}_{\mathrm{A}}=[n_{\mathrm{A},1},\ldots,n_{\mathrm{A},N_{\mathrm{I}}}]^{\mathsf{T}}\in\mathbb{C}^{N_{\mathrm{I}}\times1}$
with $n_{\mathrm{A},i}$ denoting the noise generated internally by
the $i$th amplifier (referenced to input to the amplifier) having
power $\sigma_{\mathrm{A}}^{2}=kT_{\mathrm{A}}B$ where $T_{\mathrm{A}}$
is the equivalent noise temperature of the amplifier, so that we have
$\mathbf{n}_{\mathrm{A}}\sim\mathcal{CN}(\mathbf{0},\sigma_{\mathrm{A}}^{2}\mathbf{I}_{N_{\mathrm{I}}})$. 

Substituting (\ref{eq:b-a_PA}) into (\ref{eq:a-b_RIS}), we have
that 
\begin{equation}
\mathbf{a}_{\mathrm{A}}=\left(\mathbf{I}_{N_{\mathrm{I}}}-\mathbf{\Phi}_{\mathrm{AA}}\mathbf{A}\right)^{-1}\left(\mathbf{\Phi}_{\mathrm{AI}}\left(\mathbf{b}_{\mathrm{I}}+\mathbf{n}_{\mathrm{in}}\right)+\mathbf{\Phi}_{\mathrm{AA}}\mathbf{A}\mathbf{n}_{\mathrm{A}}\right),
\end{equation}
such that 
\begin{equation}
\begin{aligned}\mathbf{b}_{\mathrm{A}}=\mathbf{A}\left(\mathbf{I}_{N_{\mathrm{I}}}-\mathbf{\Phi}_{\mathrm{AA}}\mathbf{A}\right)^{-1}\left(\mathbf{\Phi}_{\mathrm{AI}}\left(\mathbf{b}_{\mathrm{I}}+\mathbf{n}_{\mathrm{in}}\right)+\mathbf{n}_{\mathrm{A}}\right).\end{aligned}
\label{eq: ba}
\end{equation}
Substituting \eqref{eq: ba} into (\ref{eq:a-b_RIS}), we can relate
$\mathbf{a}_{\mathrm{I}}$ and $\mathbf{b}_{\mathrm{I}}$ as 
\begin{equation}
\begin{aligned}\mathbf{a}_{\mathrm{I}} & =\underbrace{\left(\mathbf{\Phi}_{\mathrm{II}}+\mathbf{\Phi}_{\mathrm{IA}}\mathbf{A}\left(\mathbf{I}_{N_{\mathrm{I}}}-\mathbf{\Phi}_{\mathrm{AA}}\mathbf{A}\right)^{-1}\mathbf{\Phi}_{\mathrm{AI}}\right)}_{=\mathbf{\Gamma}_{\mathrm{I}}}\left(\mathbf{b}_{\mathrm{I}}+\mathbf{n}_{\mathrm{in}}\right)\\
 & ~~+\underbrace{\mathbf{\Phi}_{\mathrm{IA}}\mathbf{A}\left(\mathbf{I}_{N_{\mathrm{I}}}-\mathbf{\Phi}_{\mathrm{AA}}\mathbf{A}\right)^{-1}}_{=\mathbf{\Pi}_{\mathrm{I}}}\mathbf{n}_{\mathrm{A}}.
\end{aligned}
\label{eq:ai_bi}
\end{equation}
where $\mathbf{\Gamma}_{\mathrm{I}}\in\mathbb{C}^{N_{\mathrm{I}}\times N_{\mathrm{I}}}$
denotes the scattering matrix of the $N_{\mathrm{I}}$-port active
reconfigurable impedance network, $\mathbf{\Gamma}_{\mathrm{I}}\mathbf{n}_{\mathrm{in}}+\mathbf{\Pi}_{\mathrm{I}}\mathbf{n}_{\mathrm{A}}$
denotes the dynamic noise of the $N_{\mathrm{I}}$-port active reconfigurable
impedance network, resulting from the input noise $\mathbf{n}_{\mathrm{in}}$
and the noise generated internally by the amplifiers $\mathbf{n}_{\mathrm{A}}$.

So far, the $N_{\mathrm{I}}$-port active reconfigurable impedance
network is fully characterized by (\ref{eq:ai_bi}), which however
has a intricate expression. This makes it not only hard to gain deep
insights on the active BD-RIS model, but also intractable for the
active BD-RIS optimization. Therefore, it is necessary to simplify
(\ref{eq:ai_bi}). To that end, we make the following assumptions.

\begin{assump}The $2N_{\mathrm{I}}$-port passive reconfigurable
impedance network is lossless, which yields $\mathbf{\Phi}^{\mathsf{H}}\mathbf{\Phi}=\mathbf{I}_{2N_{\mathrm{I}}}$. This can be achieved when all variable impedance components in the network are purely reactive. \end{assump}
\begin{assump}
Ports 1-$N_{\mathrm{I}}$ are perfectly matched and isolated with
each other, so are Ports $(N_{\mathrm{I}}+1)$-$2N_{\mathrm{I}}$,
which yields $\mathbf{\Phi}_{\mathrm{II}}=\mathbf{0}$ and $\mathbf{\Phi}_{\mathrm{AA}}=\mathbf{0}$. This can be achieved by properly tuning all the variable impedance components in the passive reconfigurable impedance network.
\end{assump}

When Assumptions 1 and 2 hold true, we have that 
\begin{equation}
\mathbf{\Phi}_{\mathrm{IA}}^{\mathsf{H}}\mathbf{\Phi}_{\mathrm{IA}}=\mathbf{I}_{N_{\mathrm{I}}},\:\mathbf{\Phi}_{\mathrm{AI}}^{\mathsf{H}}\mathbf{\Phi}_{\mathrm{AI}}=\mathbf{I}_{N_{\mathrm{I}}},
\end{equation}
which means that $\mathbf{\Phi}_{\mathrm{IA}}$ and $\mathbf{\Phi}_{\mathrm{AI}}$
are unitary matrices. Applying Assumption 2 and that $\mathbf{\Phi}_{\mathrm{AI}}$
is unitary, we can simplify (\ref{eq:ai_bi}) as 
\begin{align}
\mathbf{a}_{\mathrm{I}} & =\mathbf{\Phi}_{\mathrm{IA}}\mathbf{A}\mathbf{\Phi}_{\mathrm{AI}}\left(\mathbf{b}_{\mathrm{I}}+\mathbf{n}_{\mathrm{in}}+\mathbf{\Phi}_{\mathrm{AI}}^{\mathsf{H}}\mathbf{n}_{\mathrm{A}}\right).\label{eq: simplifed ai and bi}
\end{align}
Furthermore, we define $\mathbf{\Theta}\in\mathbb{C}^{N_{\mathrm{I}}\times N_{\mathrm{I}}}$
and $\mathbf{n}_{\mathrm{I}}\in\mathbb{C}^{N_{\mathrm{I}}\times1}$
as 
\begin{equation}
\mathbf{\Theta}=\mathbf{\Phi}_{\mathrm{IA}}\mathbf{A}\mathbf{\Phi}_{\mathrm{AI}},\label{eq: SVD}
\end{equation}
\begin{equation}
\mathbf{n}_{\mathrm{I}}=\mathbf{n}_{\mathrm{in}}+\mathbf{\Phi}_{\mathrm{AI}}^{\mathsf{H}}\mathbf{n}_{\mathrm{A}},
\end{equation}
so that we can rewrite \eqref{eq: simplifed ai and bi} as 
\begin{equation}
\mathbf{a}_{\mathrm{I}}=\mathbf{\Theta}\mathbf{b}_{\mathrm{I}}+\mathbf{\mathbf{\Theta}}\mathbf{n}_{\mathrm{I}},\label{eq:ai_bi_simplified}
\end{equation}
where $\mathbf{\Theta}$ is the scattering matrix of the $N_{\mathrm{I}}$-port
active reconfigurable impedance network and $\mathbf{\mathbf{\Theta}}\mathbf{n}_{\mathrm{I}}$
is the dynamic noise of the $N_{\mathrm{I}}$-port active reconfigurable
impedance network with $\mathbf{n}_{\mathrm{I}}\sim\mathcal{CN}(\mathbf{0},\sigma_{\mathrm{I}}^{2}\mathbf{I}_{N_{\mathrm{I}}})$
and $\sigma_{\mathrm{I}}^{2}=kB\left(T_{\mathrm{in}}+T_{\mathrm{A}}\right)$
resulting from the input noise and the noise generated internally
by the amplifier. The dynamic noise thus clearly varies with amplification factor of amplifiers that are embedded in $\mathbf{\Theta}$. Overall, under Assumptions 1 and 2, $\mathbf{\Gamma}_{\mathrm{I}}$
is simplified to $\mathbf{\Theta}$ and $\mathbf{\Gamma}_{\mathrm{I}}\mathbf{n}_{\mathrm{in}}+\mathbf{\Pi}_{\mathrm{I}}\mathbf{n}_{\mathrm{A}}$
is simplified to $\mathbf{\mathbf{\Theta}}\mathbf{n}_{\mathrm{I}}$.

\subsection{Architecture Design of Active BD-RIS}

The circuit topology of the $2N_{\mathrm{I}}$-port passive reconfigurable
impedance network determines the characteristics of its scattering
matrix $\mathbf{\Phi}$ including $\mathbf{\Phi}_{\mathrm{II}}$,
$\mathbf{\Phi}_{\mathrm{IA}}$, $\mathbf{\Phi}_{\mathrm{AI}}$, and
$\mathbf{\Phi}_{\mathrm{AA}}$ and thus determines the characteristics
of the scattering matrix of the $N_{\mathrm{I}}$-port active reconfigurable
impedance network, that is $\mathbf{\Gamma}_{\mathrm{I}}$ in the
general case \eqref{eq:ai_bi} or $\mathbf{\mathbf{\Theta}}$ in the
simplified case \eqref{eq: SVD}. In this subsection, following Assumptions
1 and 2, we introduce three architectures of active BD-RIS with different
circuit topology of the $2N_{\mathrm{I}}$-port passive reconfigurable
impedance network.

As a starting point, we first consider the $2N_{\mathrm{I}}$-port
passive reconfigurable impedance network is reciprocal, which is the
common case, so that we have $\mathbf{\Phi}_{\mathrm{AI}}=\mathbf{\Phi}_{\mathrm{IA}}^{\mathsf{T}}$.

\begin{figure}
\centering{}\includegraphics[width=0.48\textwidth]{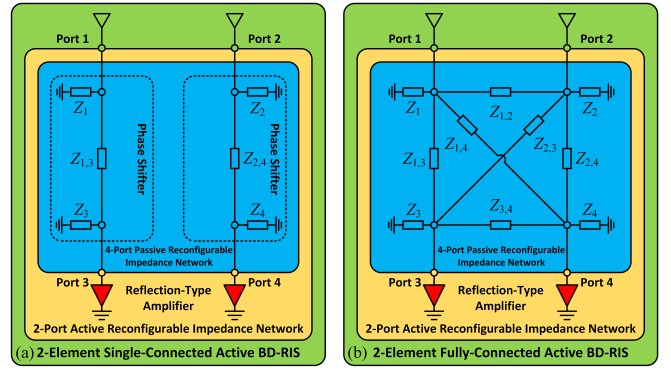}
\caption{(a) 2-element single-connected active BD-RIS (that is essentially
the 2-element active D-RIS) and (b) 2-element fully-connected active
BD-RIS.}
\label{fig:single-fully-ABDRIS}
\end{figure}

\subsubsection{Single-Connected Active BD-RIS}

For this case, the $2N_{\mathrm{I}}$-port passive reconfigurable
impedance network is constructed by $N_{\mathrm{I}}$ isolated 2-port
networks. An example for a 2-element single-connected active BD-RIS
is illustrated in Fig. \ref{fig:single-fully-ABDRIS}(a). For $i=1,\ldots,N_{\mathrm{I}}$,
the $i$th 2-port network is constructed by connecting port $i$ and
port $N_{\mathrm{I}}+i$ to ground with variable impedance $Z_{i}$
and $Z_{N_{\mathrm{I}}+i}$, respectively, and connecting port $i$
and port $N_{\mathrm{I}}+i$ through a variable impedance $Z_{i,N_{\mathrm{I}}+i}$.
All the $N_{\mathrm{I}}$ 2-port networks are not connected with each
other, so that $\mathbf{\Phi}_{\mathrm{II}}$, $\mathbf{\Phi}_{\mathrm{IA}}$,
$\mathbf{\Phi}_{\mathrm{AI}}$, and $\mathbf{\Phi}_{\mathrm{AA}}$
are all diagonal matrices. In addition, by jointly adjusting $Z_{i}$,
$Z_{N_{\mathrm{I}}+i}$, and $Z_{i,N_{\mathrm{I}}+i}$, the $i$th
2-port network can work as a phase shifter with perfect impedance
matching and variable phase shift $\phi_{i}$, that is $\left[\mathbf{\Phi}_{\mathrm{II}}\right]_{i,i}=\left[\mathbf{\Phi}_{\mathrm{AA}}\right]_{i,i}=0$
and $\left[\mathbf{\Phi}_{\mathrm{IA}}\right]_{i,i}=e^{j\phi_{i}}.$
Thus, $\mathbf{\Phi}_{\mathrm{II}}=\mathbf{\Phi}_{\mathrm{AA}}=\boldsymbol{0}$
can be satisfied and $\mathbf{\Phi}_{\mathrm{IA}}$ is a diagonal
matrix written as $\mathbf{\Phi}_{\mathrm{IA}}=\mathsf{diag}(e^{j\phi_{i}},\ldots,e^{j\phi_{N_{\mathrm{I}}}})$.
As per \eqref{eq: SVD}, the scattering matrix $\mathbf{\Theta}$
of single-connected $N_{\mathrm{I}}$-port active reconfigurable impedance
network can be any complex diagonal matrix, written as 
\begin{equation}
\mathbf{\Theta}=\mathsf{diag}(A_{1}e^{j\theta_{1}},\ldots,A_{N_{\mathrm{I}}}e^{j\theta_{N_{\mathrm{I}}}}),\label{eq:single-connected ABDRIS}
\end{equation}
where $\theta_{i}=2\phi_{i}$ is the total phase shift for port $i$
for $i=1,\ldots,N_{\mathrm{I}}$. From \eqref{eq:single-connected ABDRIS},
we can find that the single-connected active BD-RIS is essentially
same as the active D-RIS \cite{Zhang2023}.

\subsubsection{Fully-Connected Active BD-RIS}

For this case, the $2N_{\mathrm{I}}$-port passive reconfigurable
impedance network is constructed by interconnecting all ports with
variable impedance. An example for a 2-element fully-connected active
BD-RIS is illustrated in Fig. \ref{fig:single-fully-ABDRIS}(b). Generally,
for $i=1,\ldots,2N_{\mathrm{I}}$, port $i$ is connected to ground
with variable impedance $Z_{i}$ and port $i$ is also connected to
port $j$ through a variable impedance $Z_{i,j}$ for $j=i+1,\ldots,2N_{\mathrm{I}}$.
It is shown in \cite{shen2021} that when $Z_{i}$ and $Z_{i,j}$
$\forall i,j$ are purely reactive (lossless), the scattering matrix
$\mathbf{\Phi}$ of such fully-connected $2N_{\mathrm{I}}$-port passive
reconfigurable impedance network can be any complex symmetric unitary
matrix, that is $\mathbf{\Phi}=\mathbf{\Phi}^{\mathsf{T}}$ and $\mathbf{\Phi}^{\mathsf{H}}\mathbf{\Phi}=\mathbf{I}_{2N_{\mathrm{I}}}$,
so that $\mathbf{\Phi}_{\mathrm{AI}}=\mathbf{\Phi}_{\mathrm{IA}}^{\mathsf{T}}$
can be any unitary matrix while maintaining $\mathbf{\Phi}_{\mathrm{II}}=\mathbf{\Phi}_{\mathrm{AA}}=\boldsymbol{0}$.
Thus, we have $\mathbf{\Theta}=\mathbf{\Phi}_{\mathrm{IA}}\mathbf{A}\mathbf{\Phi}_{\mathrm{IA}}^{\mathsf{T}}$
as per \eqref{eq: SVD}, where $\mathbf{\Phi}_{\mathrm{IA}}$ can
be any unitary matrix and $\mathbf{A}$ can be any real non-negative
diagonal matrix. This essentially is the Takagi factorization of any
complex symmetric matrix \cite{horn2012matrix}. In other words, we
have that the scattering matrix $\mathbf{\Theta}$ of the fully-connected
$N_{\mathrm{I}}$-port active reconfigurable impedance network can
be any complex symmetric matrix, so that the corresponding constraint
is 
\begin{equation}
\mathbf{\mathbf{\Theta}}=\mathbf{\Theta}^{\mathsf{T}},\label{eq:fully-connected ABDRIS}
\end{equation}
which is more general than the single-connected active BD-RIS (that
is active D-RIS).

\subsubsection{Group-Connected Active BD-RIS}

\begin{figure}
\centering{}\includegraphics[width=0.48\textwidth]{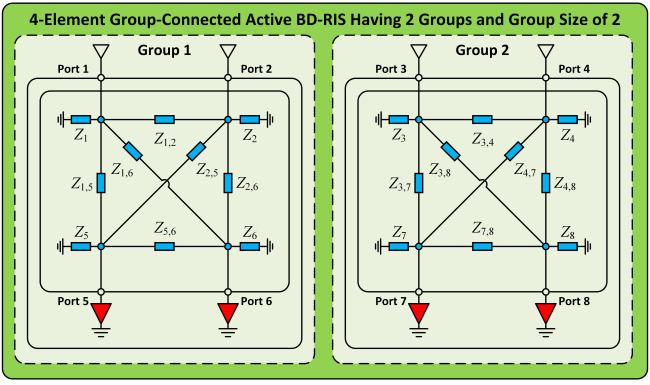}
\caption{4-element group-connected active BD-RIS having 2 groups and group
size of 2.}
\label{fig:Group-ABDRIS}
\end{figure}

The number of variable impedance in the fully-connected $2N_{\mathrm{I}}$-port
passive reconfigurable impedance network increases quadratically with
$N_{\mathrm{I}}$, which results in a high circuit complexity when
$N_{\mathrm{I}}$ is large. As a compromise, we propose the group-connected
active BD-RIS to balance the circuit complexity and performance. An
example for a 4-element group-connected active BD-RIS is illustrated
in Fig. \ref{fig:Group-ABDRIS}, where the 4 elements are divided
into 2 groups with each group having 2 elements and using the fully-connection.
Generally, we divide the $N_{\mathrm{I}}$ elements into $G$ groups
and each group has $N_{\mathrm{G}}=\frac{N_{\mathrm{I}}}{G}$ elements
and uses a fully-connected $2N_{\mathrm{G}}$-port passive reconfigurable
impedance network, where $N_{\mathrm{G}}$ refers to the group size.
It is straightforward to show that $\mathbf{\Phi}_{\mathrm{AI}}=\mathbf{\Phi}_{\mathrm{IA}}^{\mathsf{T}}$
and $\mathbf{\Phi}_{\mathrm{IA}}$ can be any block diagonal matrix
where each block is unitary, written as $\mathbf{\Phi}_{\mathrm{IA}}=\mathsf{blkdiag}(\mathbf{\Phi}_{\mathrm{IA},1},\mathbf{\Phi}_{\mathrm{IA},2},\ldots,\mathbf{\Phi}_{\mathrm{IA},G})$
with $\mathbf{\Phi}_{\mathrm{IA},g}\in\mathbb{C}^{N_{\mathrm{G}}\times N_{\mathrm{G}}}$
satisfying $\mathbf{\Phi}_{\mathrm{IA},g}^{\mathsf{H}}\mathbf{\Phi}_{\mathrm{IA},g}=\mathbf{I}_{N_{\mathrm{G}}}$.
Accordingly, we partition the diagonal matrix $\mathbf{A}$ to rewrite
it as $\mathbf{A}=\mathsf{blkdiag}(\mathbf{A}_{1},\mathbf{A}_{2},\ldots,\mathbf{A}_{G})$
with $\mathbf{A}_{g}\in\mathbb{R}^{N_{\mathrm{G}}\times N_{\mathrm{G}}}$
being a real non-negative diagonal matrix. Therefore, we have that
$\mathbf{\Theta}=\mathsf{blkdiag}(\mathbf{\Theta}_{1},\mathbf{\Theta}_{2},\ldots,\mathbf{\Theta}_{G})$
with $\mathbf{\mathbf{\Theta}}_{g}=\mathbf{\Phi}_{\mathrm{IA},g}\mathbf{A}_{g}\mathbf{\Phi}_{\mathrm{IA},g}^{\mathsf{T}}$
as per \eqref{eq: SVD}. Similarly, using the Takagi factorization,
$\mathbf{\mathbf{\Theta}}_{g}\in\mathbb{C}^{N_{\mathrm{G}}\times N_{\mathrm{G}}}$
can be any complex symmetric matrix. Thus, we have that the scattering
matrix $\mathbf{\Theta}$ of the group-connected $N_{\mathrm{I}}$-port
active reconfigurable impedance network can be any block diagonal
matrix where each block is complex symmetric, so that the corresponding
constraint is 
\begin{equation}
\mathbf{\Theta}=\mathsf{blkdiag}(\mathbf{\Theta}_{1},\mathbf{\Theta}_{2},\ldots,\mathbf{\Theta}_{G}),\mathbf{\mathbf{\Theta}}_{g}=\mathbf{\Theta}_{g}^{\mathsf{T}},\forall g.\label{eq:group-connected ABDRIS}
\end{equation}
The single- and fully-connected active BD-RISs can be particular cases
of the group-connected active BD-RIS when $N_{\mathrm{G}}=1$ and
$N_{\mathrm{G}}=N_{\mathrm{I}}$, respectively.

Next, we consider that the 2$N_{\mathrm{I}}$-port passive reconfigurable
impedance network is non-reciprocal. The analysis and modeling are
similar but based on that $\mathbf{\Phi}_{\mathrm{AI}}$ and $\mathbf{\Phi}_{\mathrm{IA}}$
are two independent unitary matrices. For brevity, we have
\begin{itemize}
\item Non-reciprocal single-connected active BD-RIS: $\mathbf{\Phi}_{\mathrm{IA}}$
and $\mathbf{\Phi}_{\mathrm{AI}}$ are diagonal matrices with the
$(i,i)$th entry being $e^{j\phi_{i}^{\mathrm{IA}}}$ and $e^{j\phi_{i}^{\mathrm{AI}}}$,
respectively, so that we have $\mathbf{\Theta}=\mathsf{diag}(A_{1}e^{j\theta_{1}},\ldots,A_{N_{\mathrm{I}}}e^{j\theta_{N_{\mathrm{I}}}})$
with $\theta_{i}=\phi_{i}^{\mathrm{IA}}+\phi_{i}^{\mathrm{AI}}$,
which is essentially same as the reciprocal case.
\item Non-reciprocal fully-connected active BD-RIS: $\mathbf{\Phi}_{\mathrm{IA}}$
and $\mathbf{\Phi}_{\mathrm{AI}}$ are independent unitary matrices,
so that we have $\mathbf{\Theta}=\mathbf{\Phi}_{\mathrm{IA}}\mathbf{A}\mathbf{\Phi}_{\mathrm{AI}}$
can be any matrix using the singular value decomposition (SVD), that
is no constraint for $\mathbf{\Theta}$.
\item Non-reciprocal group-connected active BD-RIS: $\mathbf{\Phi}_{\mathrm{IA}}$
and $\mathbf{\Phi}_{\mathrm{AI}}$ are independent block diagonal
matrices where each block is unitary, so that we have $\mathbf{\Theta}=\mathsf{blkdiag}(\mathbf{\Theta}_{1},\mathbf{\Theta}_{2},\ldots,\mathbf{\Theta}_{G})$
with $\mathbf{\mathbf{\Theta}}_{g}=\mathbf{\Phi}_{\mathrm{IA},g}\mathbf{A}_{g}\mathbf{\Phi}_{\mathrm{AI},g}$
$\forall g$ being any matrix using SVD, that is no constraint for
$\mathbf{\mathbf{\Theta}}_{g}$.
\end{itemize}
In a word, the non-reciprocal active BD-RIS can be modeled by simply
removing the symmetric constraint in the reciprocal active BD-RIS.

Comparisons of the single-, fully-, and group-connected active BD-RIS
will be provided in the following sections.

\subsection{Relationship to Passive BD-RIS}

When we replace the $N_{\mathrm{I}}$ reflection-type amplifiers with
any $N_{\mathrm{I}}$ fixed reactive loads, the $N_{\mathrm{I}}$-port
active reconfigurable impedance network becomes an $N_{\mathrm{I}}$-port
passive reconfigurable impedance network and accordingly the $N_{\mathrm{I}}$-element
active BD-RIS becomes an $N_{\mathrm{I}}$-element passive BD-RIS.
Specifically, replacing the amplifiers with fixed reactive loads leads
to $\mathbf{n}_{\mathrm{A}}=\mathbf{0}$ and $\mathbf{A}=\mathsf{diag}(e^{j\psi_{1}},\ldots,e^{j\psi_{N_{\mathrm{I}}}})$
where $e^{j\psi_{i}}$ denotes the reflection coefficient of the $i$th
fixed reactive loads, so that further ignoring the input noise $\mathbf{n}_{\mathrm{in}}$
we can transform (\ref{eq:ai_bi_simplified}) to 
\begin{equation}
\mathbf{a}_{\mathrm{I}}=\mathbf{\Theta}\mathbf{b}_{\mathrm{I}},\label{eq:passive BDRIS}
\end{equation}
where $\mathbf{\Theta}$ can be regarded as the scattering matrix
of $N_{\mathrm{I}}$-port passive reconfigurable impedance network.

Furthermore, utilizing the Takagi factorization, under the condition
that $\mathbf{A}=\mathsf{diag}(e^{j\psi_{1}},\ldots,e^{j\psi_{N_{\mathrm{I}}}})$
and $\mathbf{\Phi}_{\mathrm{IA}}$ is unitary, we have that $\mathbf{\Theta}=\mathbf{\Phi}_{\mathrm{IA}}\mathbf{A}\mathbf{\Phi}_{\mathrm{IA}}^{\mathsf{T}}$
is a complex symmetric unitary matrix (i.e. $\mathbf{\Theta}=\mathbf{\Theta}^{\mathsf{T}}$
and $\mathbf{\Theta}^{\mathsf{H}}\mathbf{\Theta}=\mathbf{I}_{N_{\mathrm{I}}}$),
which corresponds to the fully-connected passive BD-RIS \cite{shen2021}.
Therefore, it is straightforward to show that the single/fully/group-connected
active BD-RIS becomes the single/fully/group-connected passive BD-RIS
when the amplifiers are replaced with fixed reactive loads, so are
the non-reciprocal cases.

\section{Active BD-RIS Aided Communication Model}

In this section, we derive the active BD-RIS aided communication model
based on the multiport network analysis.

\subsection{General Active BD-RIS Aided Communication Model}

We start with a general active BD-RIS aided communication model without
Assumptions 1 and 2. Using \eqref{eq:tx}, \eqref{eq:rx}, and \eqref{eq:ai_bi},
we can relate $\mathbf{a}$ and $\mathbf{b}$ by 
\begin{equation}
\mathbf{a}=\mathbf{b}_{\mathrm{S}}+\mathbf{\Gamma}\mathbf{b},\label{eq: a=00003D00003Dbs+gamma*b}
\end{equation}
where $\mathbf{b}_{\mathrm{S}}$ and $\mathbf{\Gamma}$ are given
as 
\begin{equation}
\mathbf{b}_{\mathrm{S}}=\left[\begin{matrix}\mathbf{b}_{\mathrm{S},\mathrm{T}}\\
\mathbf{\Gamma}_{\mathrm{I}}\mathbf{n}_{\mathrm{in}}+\mathbf{\Pi}_{\mathrm{I}}\mathbf{n}_{\mathrm{A}}\\
\mathbf{\Gamma}_{\mathrm{R}}\mathbf{n}_{\mathrm{R}}
\end{matrix}\right],~\mathbf{\Gamma}=\left[\begin{matrix}\mathbf{\Gamma}_{\mathrm{T}} & \mathbf{0} & \mathbf{0}\\
\mathbf{0} & \mathbf{\Gamma}_{\mathrm{I}} & \mathbf{0}\\
\mathbf{0} & \mathbf{0} & \mathbf{\Gamma}_{\mathrm{R}}
\end{matrix}\right].
\end{equation}
Substituting $\mathbf{b}=\mathbf{S}\mathbf{a}$ into \eqref{eq: a=00003D00003Dbs+gamma*b},
we have 
\begin{equation}
\mathbf{b}=\underbrace{(\mathbf{I}_{N}-\mathbf{S}\mathbf{\Gamma})^{-1}\mathbf{S}}_{\triangleq\mathbf{T}}\mathbf{b}_{\mathrm{S}},
\end{equation}
where $\mathbf{T}\triangleq(\mathbf{I}_{N}-\mathbf{S}\mathbf{\Gamma})^{-1}\mathbf{S}\in\mathbb{C}^{N\times N}$
can be partitioned as 
\begin{equation}
\mathbf{T}=\left[\begin{matrix}\mathbf{T}_{\mathrm{TT}} & \mathbf{T}_{\mathrm{TI}} & \mathbf{T}_{\mathrm{TR}}\\
\mathbf{T}_{\mathrm{IT}} & \mathbf{T}_{\mathrm{II}} & \mathbf{T}_{\mathrm{IR}}\\
\mathbf{T}_{\mathrm{RT}} & \mathbf{T}_{\mathrm{RI}} & \mathbf{T}_{\mathrm{RR}}
\end{matrix}\right].
\end{equation}
As a result, we can find $\mathbf{b}_{\mathrm{T}}$ and $\mathbf{b}_{\mathrm{R}}$
as 
\begin{equation}
\begin{aligned}\mathbf{b}_{\mathrm{T}} & =\mathbf{T}_{\mathrm{TT}}\mathbf{b}_{\mathrm{S},\mathrm{T}}+\mathbf{T}_{\mathrm{TI}}\left(\mathbf{\Gamma}_{\mathrm{I}}\mathbf{n}_{\mathrm{in}}+\mathbf{\Pi}_{\mathrm{I}}\mathbf{n}_{\mathrm{A}}\right)+\mathbf{T}_{\mathrm{TR}}\mathbf{\Gamma}_{\mathrm{R}}\mathbf{n}_{\mathrm{R}},\\
\mathbf{b}_{\mathrm{R}} & =\mathbf{T}_{\mathrm{RT}}\mathbf{b}_{\mathrm{S},\mathrm{T}}+\mathbf{T}_{\mathrm{RI}}\left(\mathbf{\Gamma}_{\mathrm{I}}\mathbf{n}_{\mathrm{in}}+\mathbf{\Pi}_{\mathrm{I}}\mathbf{n}_{\mathrm{A}}\right)+\mathbf{T}_{\mathrm{RR}}\mathbf{\Gamma}_{\mathrm{R}}\mathbf{n}_{\mathrm{R}}.
\end{aligned}
\label{eq:bt_br}
\end{equation}
The voltage vectors at the transmitter and receiver are denoted as
$\mathbf{v}_{\mathrm{T}}=[v_{\mathrm{T},1},\ldots,v_{\mathrm{T},N_{\mathrm{T}}}]^{\mathsf{T}}\in\mathbb{C}^{N_{\mathrm{T}}\times1}$
and $\mathbf{v}_{\mathrm{R}}=[v_{\mathrm{R},1},\ldots,v_{\mathrm{R},N_{\mathrm{R}}}]^{\mathsf{T}}\in\mathbb{C}^{N_{\mathrm{R}}\times1}$,
respectively, where $v_{\mathrm{T},n_{\mathrm{T}}}$ and $v_{\mathrm{R},n_{\mathrm{R}}}$
are the voltage at the $n_{\mathrm{T}}$th transmit antenna and the
$n_{\mathrm{R}}$th receive antenna, respectively. Accordingly, using
(\ref{eq:tx}), (\ref{eq:rx}), and (\ref{eq:bt_br}), we can express
$\mathbf{v}_{\mathrm{T}}$ and $\mathbf{v}_{\mathrm{R}}$ by 
\begin{equation}
\begin{aligned}\mathbf{v}_{\mathrm{T}} & =\mathbf{a}_{\mathrm{T}}+\mathbf{b}_{\mathrm{T}}=\mathbf{b}_{\mathrm{S},\mathrm{T}}+\left(\mathbf{I}_{N_{\mathrm{T}}}+\mathbf{\Gamma}_{\mathrm{T}}\right)\mathbf{b}_{\mathrm{T}}\\
 & =\left(\mathbf{I}_{N_{\mathrm{T}}}+\mathbf{T}_{\mathrm{TT}}+\mathbf{\Gamma}_{\mathrm{T}}\mathbf{T}_{\mathrm{TT}}\right)\mathbf{b}_{\mathrm{S},\mathrm{T}}+\left(\mathbf{I}_{N_{\mathrm{T}}}+\mathbf{\Gamma}_{\mathrm{T}}\right)\\
 & \:\:\times\left(\mathbf{T}_{\mathrm{TI}}\left(\mathbf{\Gamma}_{\mathrm{I}}\mathbf{n}_{\mathrm{in}}+\mathbf{\Pi}_{\mathrm{I}}\mathbf{n}_{\mathrm{A}}\right)+\mathbf{T}_{\mathrm{TR}}\mathbf{\Gamma}_{\mathrm{R}}\mathbf{n}_{\mathrm{R}}\right),\\
 & \overset{\text{(a)}}{=}\left(\mathbf{I}_{N_{\mathrm{T}}}+\mathbf{T}_{\mathrm{TT}}+\mathbf{\Gamma}_{\mathrm{T}}\mathbf{T}_{\mathrm{TT}}\right)\mathbf{b}_{\mathrm{S},\mathrm{T}},
\end{aligned}
\label{eq:vt}
\end{equation}
where (a) holds true since the strength of the noise is negligible
compared to the strength of signal at the transmitter, and that 
\begin{equation}
\begin{aligned}\mathbf{v}_{\mathrm{R}} & =\mathbf{a}_{\mathrm{R}}+\mathbf{b}_{\mathrm{R}}+\mathbf{n}_{\mathrm{R}}=\left(\mathbf{I}_{N_{\mathrm{R}}}+\mathbf{\Gamma}_{\mathrm{R}}\right)(\mathbf{b}_{\mathrm{R}}+\mathbf{n}_{\mathrm{R}})\\
 & =\left(\mathbf{I}_{N_{\mathrm{R}}}+\mathbf{\Gamma}_{\mathrm{R}}\right)(\mathbf{T}_{\mathrm{RT}}\mathbf{b}_{\mathrm{S},\mathrm{T}}+\mathbf{T}_{\mathrm{RI}}\left(\mathbf{\Gamma}_{\mathrm{I}}\mathbf{n}_{\mathrm{in}}+\mathbf{\Pi}_{\mathrm{I}}\mathbf{n}_{\mathrm{A}}\right))\\
 & \:\:+\left(\mathbf{I}_{N_{\mathrm{R}}}+\mathbf{\Gamma}_{\mathrm{R}}\right)\left(\mathbf{I}_{N_{\mathrm{R}}}+\mathbf{T}_{\mathrm{RR}}\mathbf{\Gamma}_{\mathrm{R}}\right)\mathbf{n}_{\mathrm{R}}.
\end{aligned}
\end{equation}
We assign $\mathbf{v}_{\mathrm{T}}$ and $\mathbf{v}_{\mathrm{R}}$
as the transmit signal and receive signal, respectively, so that by
establishing the relationship between $\mathbf{v}_{\mathrm{T}}$ and
$\mathbf{v}_{\mathrm{R}}$ we can find the active BD-RIS aided communication
model, written as 
\begin{equation}
\begin{aligned}\mathbf{v}_{\mathrm{R}} & =\underbrace{\left(\mathbf{I}_{N_{\mathrm{R}}}+\mathbf{\Gamma}_{\mathrm{R}}\right)\mathbf{T}_{\mathrm{RT}}\left(\mathbf{I}_{N_{\mathrm{T}}}+\mathbf{T}_{\mathrm{TT}}+\mathbf{\Gamma}_{\mathrm{T}}\mathbf{T}_{\mathrm{TT}}\right)^{-1}}_{=\mathbf{H}}\mathbf{v}_{\mathrm{T}}\\
 & \:\:+\left(\mathbf{I}_{N_{\mathrm{R}}}+\mathbf{\Gamma}_{\mathrm{R}}\right)\mathbf{T}_{\mathrm{RI}}\left(\mathbf{\Gamma}_{\mathrm{I}}\mathbf{n}_{\mathrm{in}}+\mathbf{\Pi}_{\mathrm{I}}\mathbf{n}_{\mathrm{A}}\right)\\
 & \:\:+\left(\mathbf{I}_{N_{\mathrm{R}}}+\mathbf{\Gamma}_{\mathrm{R}}\right)(\mathbf{I}_{N_{\mathrm{R}}}+\mathbf{T}_{\mathrm{RR}}\mathbf{\Gamma}_{\mathrm{R}})\mathbf{n}_{\mathrm{R}},
\end{aligned}
\label{eq:channel}
\end{equation}
where $\mathbf{H}\in\mathbb{C}^{N_{\mathrm{R}}\times N_{\mathrm{T}}}$
denotes the channel matrix for the active BD-RIS aided communication.

The general active BD-RIS aided communication model (\ref{eq:channel})
is beneficial in terms of characterizing the general $N_{\mathrm{I}}$-port
active reconfigurable impedance network of active BD-RIS, impedance
mismatching, and mutual coupling of antennas. Nevertheless, the expression
of channel matrix $\mathbf{H}$ is complicated because of the matrix
inversion operation, which makes it hard to gain engineering insights
and optimize the active BD-RIS. To address the concern, we develop
a simplified active BD-RIS aided communication model in the following.

\subsection{Simplified Active BD-RIS Aided Communication Model}

To simplify the active BD-RIS aided communication model in (\ref{eq:channel}),
with Assumptions 1 and 2, we make the following additional assumptions.
\begin{assump} The source impedance at the transmitter and the load
impedance at the receiver are equal to $Z_{0}$, i.e. $Z_{\mathrm{T},1}=\ldots Z_{\mathrm{T},N_{\mathrm{T}}}=Z_{0}$
and $Z_{\mathrm{R},1}=\ldots=Z_{\mathrm{R},N_{\mathrm{R}}}=Z_{0}$.
Then we have $\mathbf{\Gamma}_{\mathrm{T}}=\mathbf{0}$ and $\mathbf{\Gamma}_{\mathrm{R}}=\mathbf{0}$.
\end{assump} \begin{assump} Unilateral approximation \cite{ivrlavc2010toward},
i.e. the electrical properties at the transmitter are approximately
independent of those at the receiver when large transmission distances
are assumed between devices. This corresponds to the general far-field communication system and yields $\mathbf{S}_{\mathrm{TR}}=\mathbf{0}$,
$\mathbf{S}_{\mathrm{IR}}=\mathbf{0}$, and $\mathbf{S}_{\mathrm{TI}}=\mathbf{0}$. 
\end{assump} 
\begin{assump} The antennas at transmitter, active
BD-RIS, and receiver have perfect matching with no mutual coupling.
This yields $\mathbf{S}_{\mathrm{TT}}=\mathbf{0}$, $\mathbf{S}_{\mathrm{II}}=\mathbf{0}$,
and $\mathbf{S}_{\mathrm{RR}}=\mathbf{0}$. This can be approximately achieved by individually matching each antenna to the reference impedance $Z_0$ and keeping the antenna spacing larger than half-wavelength. \end{assump} Assumptions
3, 4, and 5 yield that $\mathbf{T}$ has only three nonzero blocks, written
as 
\begin{equation}
\mathbf{T}_{\mathrm{IT}}=\mathbf{S}_{\mathrm{IT}},~\mathbf{T}_{\mathrm{RT}}=\mathbf{S}_{\mathrm{RT}}+\mathbf{S}_{\mathrm{RI}}\mathbf{\Gamma}_{\mathrm{I}}\mathbf{S}_{\mathrm{IT}},~\mathbf{T}_{\mathrm{RI}}=\mathbf{S}_{\mathrm{RI}}.
\end{equation}
Accordingly, together with Assumptions 1 and 2 which yield that $\mathbf{\Gamma}_{\mathrm{I}}$
is simplified to $\mathbf{\Theta}$ and $\mathbf{\Gamma}_{\mathrm{I}}\mathbf{n}_{\mathrm{in}}+\mathbf{\Pi}_{\mathrm{I}}\mathbf{n}_{\mathrm{A}}$
is simplified to $\mathbf{\mathbf{\Theta}}\mathbf{n}_{\mathrm{I}}$,
we can simplify (\ref{eq:channel}) as 
\begin{equation}
\mathbf{v}_{\mathrm{R}}=\underbrace{(\mathbf{S}_{\mathrm{RT}}+\mathbf{S}_{\mathrm{RI}}\mathbf{\Theta}\mathbf{S}_{\mathrm{IT}})}_{=\mathbf{H}}\mathbf{v}_{\mathrm{T}}+\mathbf{S}_{\mathrm{RI}}\mathbf{\Theta}\mathbf{n}_{\mathrm{I}}+\mathbf{n}_{\mathrm{R}}.\label{eq:channel1}
\end{equation}
 To facilitate understanding, we introduce auxiliary notations $\mathbf{H}_{\mathrm{RT}}=\mathbf{S}_{\mathrm{RT}}$,
$\mathbf{H}_{\mathrm{IT}}=\mathbf{S}_{\mathrm{IT}}$, $\mathbf{H}_{\mathrm{RI}}=\mathbf{S}_{\mathrm{RI}}$,
$\mathbf{x}=\mathbf{v}_{\mathrm{T}}$, and $\mathbf{y}=\mathbf{v}_{\mathrm{R}}$,
so that we can rewrite the simplified active BD-RIS aided communication
model (\ref{eq:channel1}) as 
\begin{equation}
\mathbf{y}=(\mathbf{H}_{\mathrm{RT}}+\mathbf{H}_{\mathrm{RI}}\mathbf{\Theta}\mathbf{H}_{\mathrm{IT}})\mathbf{x}+\mathbf{H}_{\mathrm{RI}}\mathbf{\Theta}\mathbf{n}_{\mathrm{I}}+\mathbf{n}_{\mathrm{R}},\label{eq:channel2}
\end{equation}
where $\mathbf{\Theta}\mathbf{n}_{\mathrm{I}}$ is the dynamic noise
of the active BD-RIS, $\mathbf{n}_{\mathrm{R}}$ is the noise at the
receiver, and $\mathbf{\Theta}$ has different constraints \eqref{eq:single-connected ABDRIS},
\eqref{eq:fully-connected ABDRIS}, and \eqref{eq:group-connected ABDRIS}
depending on the single/fully/group-connected active BD-RIS.

From the simplified active BD-RIS aided communication model \eqref{eq:channel2},
we can find that the channel matrix and dynamic noise are linear functions
of $\mathbf{\Theta}$, which makes the active BD-RIS optimization
tractable. It should be noted that the conventional active D-RIS aided
communication model \cite{Zhang2023} is a special case of \eqref{eq:channel2}
when the single-connected active BD-RIS is considered, i.e. $\mathbf{\Theta}=\mathsf{diag}(A_{1}e^{j\theta_{1}},\ldots,A_{N_{\mathrm{I}}}e^{j\theta_{N_{\mathrm{I}}}})$.

\section{Active BD-RIS Aided SISO System}

To gain insights into the fundamental benefits of active BD-RIS over
active D-RIS and passive BD-RIS, we analyze the received SNR of a
SISO system aided by an $N_{\mathrm{I}}$-element active BD-RIS. The
channels for such a system include those from the transmitter to receiver,
from the transmitter to active BD-RIS, and from the active BD-RIS
to receiver respectively given by $h_{\mathrm{RT}}\in\mathbb{C}$,
$\mathbf{h}_{\mathrm{IT}}\in\mathbb{C}^{N_{\mathrm{I}}\times1}$,
and $\mathbf{h}_{\mathrm{RI}}\in\mathbb{C}^{1\times N_{\mathrm{I}}}$.
Let $x\in\mathbb{C}$ denote the transmit signal with $x=\sqrt{P_{\mathrm{T}}}s$,
where $s\in\mathbb{C}$ denotes the transmit symbol with $\mathbb{E}\{|s|^{2}\}=1$
and $P_{\mathrm{T}}$ denotes the transmit power. The received signal
at the user is given by 
\begin{equation}
y=\sqrt{P_{\mathrm{T}}}(h_{\mathrm{RT}}+\mathbf{h}_{\mathrm{RI}}\mathbf{\Theta}\mathbf{h}_{\mathrm{IT}})s+\mathbf{h}_{\mathrm{RI}}\mathbf{\Theta}\mathbf{n}_{\mathrm{I}}+n_{\mathrm{R}},
\end{equation}
where $n_{\mathrm{R}}\sim\mathcal{CN}(0,\sigma_{\mathrm{R}}^{2})$.
The SNR is thus given by 
\begin{equation}
\gamma(\mathbf{\Theta})=\frac{P_{\mathrm{T}}|h_{\mathrm{RT}}+\mathbf{h}_{\mathrm{RI}}\mathbf{\Theta}\mathbf{h}_{\mathrm{IT}}|^{2}}{\sigma_{\mathrm{I}}^{2}\|\mathbf{h}_{\mathrm{RI}}\mathbf{\Theta}\|_{2}^{2}+\sigma_{\mathrm{R}}^{2}}.
\end{equation}
In the active BD-RIS aided SISO system, the radiated power $P_\mathrm{rad}$ of active BD-RIS is constrained by\footnote{The power consumption of an active BD-RIS with $N_\mathrm{I}$ amplifiers and $N_\mathrm{re}$ variable impedance components is modeled as
$P_\mathrm{RIS}^\mathrm{tot} = N_\mathrm{re}P_\mathrm{cir} + N_\mathrm{I}P_\mathrm{DC} + \eta^{-1}P_\mathrm{rad}$,
where $P_\mathrm{cir}$ denotes the switch and control circuit power consumption for each variable impedance component, $P_\mathrm{DC}$ denotes the DC biasing power consumption for each amplifier, and $\eta$ denotes the amplifier efficiency \cite{Long2021}. The radiated power $P_\mathrm{rad}$ is limited due to the overall power budget at the active BD-RIS. Therefore, we consider the constraint $P_\mathrm{rad} \le \eta(P_\mathrm{RIS}^\mathrm{tot} - N_\mathrm{re}P_\mathrm{cir} - N_\mathrm{I}P_\mathrm{DC}) = P_\mathrm{A}$. This implies that $P_\mathrm{A}$ is essentially a function of the amplifier DC power and circuit power.} 
\begin{equation}
    \begin{aligned} 
        P_\mathrm{rad} &= \mathbb{E}\{\|\sqrt{P_{\mathrm{T}}}\mathbf{\Theta}\mathbf{h}_{\mathrm{IT}}s+\mathbf{\Theta}\mathbf{n}_{\mathrm{I}}\|_{2}^{2}\}\\
        &=P_{\mathrm{T}}\|\mathbf{\Theta}\mathbf{h}_{\mathrm{IT}}\|_{2}^{2}+\sigma_{\mathrm{I}}^{2}\|\mathbf{\Theta}\|_{\mathsf{F}}^{2}\le P_\mathrm{A},
    \end{aligned}
\end{equation} 
where $P_\mathrm{A}$ is the radiated power budget of active BD-RIS.


Then, the SNR maximization problem can be formulated as 
\begin{subequations}
\label{eq:prob_siso} 
\begin{align}
\max_{\mathbf{\Theta}}~ & \frac{P_{\mathrm{T}}|h_{\mathrm{RT}}+\mathbf{h}_{\mathrm{RI}}\mathbf{\Theta}\mathbf{h}_{\mathrm{IT}}|^{2}}{\sigma_{\mathrm{I}}^{2}\|\mathbf{h}_{\mathrm{RI}}\mathbf{\Theta}\|_{2}^{2}+\sigma_{\mathrm{R}}^{2}}\\
\mathrm{s.t.}~ & P_{\mathrm{T}}\|\mathbf{\Theta}\mathbf{h}_{\mathrm{IT}}\|_{2}^{2}+\sigma_{\mathrm{I}}^{2}\|\mathbf{\Theta}\|_{\mathsf{F}}^{2}\le P_{\mathrm{A}},\label{eq:constraint_theta1}\\
 & \mathbf{\Theta}=\mathsf{blkdiag}(\mathbf{\Theta}_{1},\ldots,\mathbf{\Theta}_{G}),\mathbf{\Theta}_{g}\in\mathcal{T}_{g},\forall g,\label{eq:constraint_theta2}
\end{align}
\end{subequations}
 where $\mathcal{T}_{g}=\{\mathbf{\Theta}_{g}~|~\mathbf{\Theta}_{g}\in\mathbb{C}^{N_{\mathrm{G}}\times N_{\mathrm{G}}}\}$,
$\forall g$, for non-reciprocal active BD-RIS and $\mathcal{T}_{g}=\{\mathbf{\Theta}_{g}~|~\mathbf{\Theta}_{g}\in\mathbb{C}^{N_{\mathrm{G}}\times N_{\mathrm{G}}},\mathbf{\Theta}_{g}=\mathbf{\Theta}_{g}^{\mathsf{T}}\}$,
$\forall g$, for reciprocal active BD-RIS. Note that we focus on
the group-connected architecture since it is a general expression
including single- and fully-connected active BD-RIS as two special
examples.

\subsection{Simplified Case with $A_{1}=\ldots=A_{N_{\mathrm{I}}}=A$ and $h_{\mathrm{RT}}=0$}

For simplicity, we assume the direct link from the transmitter and
the receiver is blocked, i.e. $h_{\mathrm{RT}}=0$, and the amplification
factors for all power amplifiers are identical, i.e. $A_{1}=\ldots=A_{N_{\mathrm{I}}}= A$.
Note that the matrix $\mathbf{A}=\mathsf{diag}(A_{1},\ldots,A_{N_{\mathrm{I}}})$
is always diagonal regardless of architectures of the passive reconfigurable impedance network. Then we have $\mathbf{\Theta} = \mathbf{\Phi}_\mathrm{IA}\mathbf{A}\mathbf{\Phi}_\mathrm{AI} = A\bar{\mathbf{\Theta}}$ where $\bar{\mathbf{\Theta}} =\mathbf{\Phi}_{\mathrm{IA}}\mathbf{\Phi}_{\mathrm{AI}}=\mathsf{blkdiag}(\bar{\mathbf{\Theta}}_{1},\ldots,\bar{\mathbf{\Theta}}_{G})$
is a block-diagonal matrix with each block being unitary, i.e. $\bar{\mathbf{\Theta}}_{g}^{\mathsf{H}}\bar{\mathbf{\Theta}}_{g}=\mathbf{I}_{N_{\mathrm{G}}}$, $\forall g$. The SNR is thus simplified to 
\begin{equation}
\gamma(\bar{\mathbf{\Theta}},A)=\frac{P_{\mathrm{T}}A^{2}|\mathbf{h}_{\mathrm{RI}}\bar{\mathbf{\Theta}}\mathbf{h}_{\mathrm{IT}}|^{2}}{\sigma_{\mathrm{I}}^{2}A^{2}\|\mathbf{h}_{\mathrm{RI}}\|_{2}^{2}+\sigma_{\mathrm{R}}^{2}}.\label{eq:snr_simplified}
\end{equation}

From (\ref{eq:snr_simplified}) we observe that the SNR is proportional
to the value of $A$. To maximize the SNR, the reflected power of
active BD-RIS is given by 
\begin{equation}
P_{\mathrm{T}}A^{2}\|\mathbf{h}_{\mathrm{IT}}\|_{\mathsf{2}}^{2}+\sigma_{\mathrm{I}}^{2}A^{2}N_{\mathrm{I}}=P_{\mathrm{A}}.
\end{equation}
Therefore, we have 
\begin{equation}
A=\sqrt{\frac{P_{\mathrm{A}}}{P_{\mathrm{T}}\|\mathbf{h}_{\mathrm{IT}}\|_{2}^{2}+\sigma_{\mathrm{I}}^{2}N_{\mathrm{I}}}},\label{eq:A}
\end{equation}
such that the SNR is rewritten as 
\begin{equation}
\bar{\gamma}(\bar{\mathbf{\Theta}})=\frac{P_{\mathrm{T}}P_{\mathrm{A}}|\mathbf{h}_{\mathrm{RI}}\bar{\mathbf{\Theta}}\mathbf{h}_{\mathrm{IT}}|^{2}}{\sigma_{\mathrm{I}}^{2}P_{\mathrm{A}}\|\mathbf{h}_{\mathrm{RI}}\|_{2}^{2}+\sigma_{\mathrm{R}}^{2}(P_{\mathrm{T}}\|\mathbf{h}_{\mathrm{IT}}\|_{2}^{2}+\sigma_{\mathrm{I}}^{2}N_{\mathrm{I}})}.
\end{equation}
For active BD-RIS with generally $\bar{\mathbf{\Theta}}=\mathsf{diag}(\bar{\mathbf{\Theta}}_{1},\ldots,\bar{\mathbf{\Theta}}_{G})$
and $\bar{\mathbf{\Theta}}_{g}^{\mathsf{H}}\bar{\mathbf{\Theta}}_{g}=\mathbf{I}_{N_{\mathrm{G}}}$,
$\forall g$, we have that $|\mathbf{h}_{\mathrm{RI}}\bar{\mathbf{\Theta}}\mathbf{h}_{\mathrm{IT}}|^{2}=|\sum_{g=1}^{G}\mathbf{h}_{\mathrm{RI},g}\bar{\mathbf{\Theta}}_{g}\mathbf{h}_{\mathrm{IT},g}|^{2}$,
where $\mathbf{h}_{\mathrm{RI},g}=[\mathbf{h}_{\mathrm{RI}}]_{(g-1)N_{\mathrm{G}}+1:gN_{\mathrm{G}}}$
and $\mathbf{h}_{\mathrm{IT},g}=[\mathbf{h}_{\mathrm{IT}}]_{(g-1)N_{\mathrm{G}}+1:gN_{\mathrm{G}}}$,
$\forall g$. Applying the Cauchy-Schwarz inequality, we have $|\sum_{g=1}^{G}\mathbf{h}_{\mathrm{RI},g}\bar{\mathbf{\Theta}}_{g}\mathbf{h}_{\mathrm{IT},g}|^{2}\le(\sum_{g=1}^{G}\|\mathbf{h}_{\mathrm{RI},g}\|_{2}\|\mathbf{h}_{\mathrm{IT},g}\|_{2})^{2}$,
where the equality holds when $\frac{\mathbf{h}_{\mathrm{RI},g}^{\mathsf{H}}}{\|\mathbf{h}_{\mathrm{RI},g}\|_{2}}=\bar{\mathbf{\Theta}}_{g}\frac{\mathbf{h}_{\mathrm{IT},g}}{\|\mathbf{h}_{\mathrm{IT},g}\|_{2}}$,
$\forall g$. For the case of a non-reciprocal active BD-RIS, we can
have a closed-form solution 
\begin{equation}
\bar{\mathbf{\Theta}}_{g}=\mathbf{V}_{\mathrm{RI},g}\mathbf{U}_{\mathrm{IT},g}^{\mathsf{H}},\forall g,\label{eq:BD_RIS}
\end{equation}
constructed by two unitary matrices $\mathbf{V}_{\mathrm{RI},g}\in\mathbb{C}^{N_{\mathrm{G}}\times N_{\mathrm{G}}}$
and $\mathbf{U}_{\mathrm{IT},g}\in\mathbb{C}^{N_{\mathrm{G}}\times N_{\mathrm{G}}}$,
where their first column are respectively given by $[\mathbf{V}_{\mathrm{RI},g}]_{:,1}=\frac{\mathbf{h}_{\mathrm{RI},g}^{\mathsf{H}}}{\|\mathbf{h}_{\mathrm{RI},g}\|_{2}}$
and $[\mathbf{U}_{\mathrm{IT},g}]_{:,1}=\frac{\mathbf{h}_{\mathrm{IT},g}}{\|\mathbf{h}_{\mathrm{IT},g}\|_{2}}$,
$\forall g$. For the case of a reciprocal active BD-RIS, we have
\begin{equation}
\bar{\mathbf{\Theta}}_{g}^{\mathsf{H}}\bar{\mathbf{\Theta}}_{g}=\mathbf{I}_{N_{\mathrm{G}}},~\bar{\mathbf{\Theta}}_{g}=\bar{\mathbf{\Theta}}_{g}^{\mathsf{T}}.
\end{equation}
We can thus apply the closed-form globally optimal solution for passive
BD-RIS proposed in \cite{nerini2023closed,santamaria2023snr} to directly
find such $\bar{\mathbf{\Theta}}_{g}$, $\forall g$. This allows
us to derive the maximum SNR achieved by active BD-RIS as 
\begin{equation}
\bar{\gamma}^{\mathsf{active,BD}}=\frac{P_{\mathrm{T}}P_{\mathrm{A}}(\sum_{g=1}^{G}\|\mathbf{h}_{\mathrm{RI},g}\|_{2}\|\mathbf{h}_{\mathrm{IT},g}\|_{2})^{2}}{\sigma_{\mathrm{I}}^{2}P_{\mathrm{A}}\|\mathbf{h}_{\mathrm{RI}}\|_{2}^{2}+\sigma_{\mathrm{R}}^{2}(P_{\mathrm{T}}\|\mathbf{h}_{\mathrm{IT}}\|_{2}^{2}+\sigma_{\mathrm{I}}^{2}N_{\mathrm{I}})}.
\end{equation}
Below, we will derive the SNR scaling law w.r.t. $N_{\mathrm{I}}$,
assuming independently distributed Rayleigh fading channels.

\subsection{SNR Scaling Law}

\subsubsection{SNR Scaling Law for Active BD-RIS}

Assuming two channels are distributed as $\mathbf{h}_{\mathrm{RI}}\sim\mathcal{CN}(\mathbf{0},\zeta_{\mathrm{RI}}^{2}\mathbf{I}_{N_{\mathrm{I}}})$
and $\mathbf{h}_{\mathrm{IT}}\sim\mathcal{CN}(\mathbf{0},\zeta_{\mathrm{IT}}^{2}\mathbf{I}_{N_{\mathrm{I}}})$,
we have $\mathbb{E}\{\|\mathbf{h}_{\mathrm{RI},g}\|_{2}\}=\frac{\Gamma(N_{\mathrm{G}}+\frac{1}{2})}{\Gamma(N_{\mathrm{G}})}\zeta_{\mathrm{RI}}$
and $\mathbb{E}\{\|\mathbf{h}_{\mathrm{IT},g}\|_{2}\}=\frac{\Gamma(N_{\mathrm{G}}+\frac{1}{2})}{\Gamma(N_{\mathrm{G}})}\zeta_{\mathrm{IT}}$,
$\forall g$, where $\Gamma(\cdot)$ is the gamma function and $\Gamma(n+\frac{1}{2})=\frac{(2n)!}{4^{n}n!}\sqrt{\pi}$
for integer $n$. In addition, we have $\mathbb{E}\{\|\mathbf{h}_{\mathrm{RI}}\|_{2}^{2}\}=N_{\mathrm{I}}\zeta_{\mathrm{RI}}^{2}$
and $\mathbb{E}\{\|\mathbf{h}_{\mathrm{IT}}\|_{2}^{2}\}=N_{\mathrm{I}}\zeta_{\mathrm{IT}}^{2}$.
Therefore, the asymptotic SNR for $N_{\mathrm{I}}\rightarrow\infty$
is derived as 
\begin{equation}
\bar{\gamma}^{\mathsf{active,BD}}\overset{N_{\mathrm{I}}\nearrow}{\approx}\alpha\frac{N_{\mathrm{I}}\Gamma^{4}(N_{\mathrm{G}}+\frac{1}{2})}{N_{\mathrm{G}}^{2}\Gamma^{4}(N_{\mathrm{G}})},\label{eq:asymp_snr_BD_gen}
\end{equation}
where $\alpha=\frac{P_{\mathrm{T}}P_{\mathrm{A}}\zeta_{\mathrm{RI}}^{2}\zeta_{\mathrm{IT}}^{2}}{\sigma_{\mathrm{I}}^{2}P_{\mathrm{A}}\zeta_{\mathrm{RI}}^{2}+\sigma_{\mathrm{R}}^{2}P_{\mathrm{T}}\zeta_{\mathrm{IT}}^{2}+\sigma_{\mathrm{R}}^{2}\sigma_{\mathrm{I}}^{2}}$.

Specifically for active BD-RIS with $G=1$, i.e. fully-connected,
the maximum SNR is given by 
\begin{equation}
\tilde{\bar{\gamma}}^{\mathsf{active,BD}}=\frac{P_{\mathrm{T}}P_{\mathrm{A}}\|\mathbf{h}_{\mathrm{RI}}\|_{2}^{2}\|\mathbf{h}_{\mathrm{IT}}\|_{2}^{2}}{\sigma_{\mathrm{I}}^{2}P_{\mathrm{A}}\|\mathbf{h}_{\mathrm{RI}}\|_{2}^{2}+\sigma_{\mathrm{R}}^{2}(P_{\mathrm{T}}\|\mathbf{h}_{\mathrm{IT}}\|_{2}^{2}+\sigma_{\mathrm{I}}^{2}N_{\mathrm{I}})},
\end{equation}
and the asymptotic SNR for $N_{\mathrm{I}}\rightarrow\infty$ is thus
\begin{equation}
\tilde{\bar{\gamma}}^{\mathsf{active,BD}}\overset{N_{\mathrm{I}}\nearrow}{\approx}\alpha N_{\mathrm{I}}.\label{eq:asymp_snr_BD}
\end{equation}

\subsubsection{SNR Scaling Law for Active D-RIS}

For the active D-RIS with $\bar{\mathbf{\Theta}}=\mathsf{diag}(e^{\jmath\theta_{1}},\ldots,e^{\jmath\theta_{N_{\mathrm{I}}}})$,
we have 
\begin{equation}
\theta_{n}=-\angle([\mathbf{h}_{\mathrm{RI}}]_{n}[\mathbf{h}_{\mathrm{IT}}]_{n}),\forall n=1,\ldots,N_{\mathrm{I}},\label{eq:D_RIS}
\end{equation}
to achieve the maximum SNR as 
\begin{equation}
\bar{\gamma}^{\mathsf{active,D}}=\frac{P_{\mathrm{T}}P_{\mathrm{A}}\Big(\sum_{n=1}^{N_{\mathrm{I}}}|[\mathbf{h}_{\mathrm{RI}}]_{n}||[\mathbf{h}_{\mathrm{IT}}]_{n}|\Big)^{2}}{\sigma_{\mathrm{I}}^{2}P_{\mathrm{A}}\|\mathbf{h}_{\mathrm{RI}}\|_{2}^{2}+\sigma_{\mathrm{R}}^{2}(P_{\mathrm{T}}\|\mathbf{h}_{\mathrm{IT}}\|_{2}^{2}+\sigma_{\mathrm{I}}^{2}N_{\mathrm{I}})}.
\end{equation}
By setting $N_{\mathrm{G}}=1$ in (\ref{eq:asymp_snr_BD_gen}), we
can directly obtain the asymptotic SNR as $N_{\mathrm{I}}\rightarrow\infty$
as 
\begin{equation}
\bar{\gamma}^{\mathsf{active,D}}\overset{N_{\mathrm{I}}\nearrow}{\approx}\alpha\frac{\pi^{2}}{16}N_{\mathrm{I}},\label{eq:asymp_snr_D}
\end{equation}
which is consistent with the result in \cite{Zhang2023}.

\begin{remark} When the power amplification factors for all power
amplifiers are identical, the benefit of active BD-RIS over active
D-RIS comes purely from the flexibility of lossless reconfigurable
impedance network, same as the passive case. Recall that with $N_{\mathrm{I}}\rightarrow\infty$,
the asymptotic SNR for D-RIS is given by \cite{wu2021intelligent}
\begin{equation}
\begin{aligned}\bar{\gamma}^{\mathsf{passive,D}} & =\frac{P_{\mathrm{T}}^{\mathsf{passive}}\Big(\sum_{n=1}^{N_{\mathrm{I}}}|[\mathbf{h}_{\mathrm{RI}}]_{n}||[\mathbf{h}_{\mathrm{IT}}]_{n}|\Big)^{2}}{\sigma_{\mathrm{R}}^{2}}\overset{N_{\mathrm{I}}\nearrow}{\approx}\beta\frac{\pi^{2}}{16}N_{\mathrm{I}}^{2},\end{aligned}
\label{eq:asymp_snr_D_passive}
\end{equation}
where $\beta=\frac{P_{\mathrm{T}}^{\mathsf{passive}}\zeta_{\mathrm{RI}}^{2}\zeta_{\mathrm{IT}}^{2}}{\sigma_{\mathrm{R}}^{2}}$
with $P_{\mathrm{T}}^{\mathsf{passive}}$ denoting the transmit power
for a passive RIS-aided system. The asymptotic SNR for passive BD-RIS
is given by \cite{shen2021} 
\begin{equation}
\bar{\gamma}^{\mathsf{passive,BD}}=\frac{P_{\mathrm{T}}^{\mathsf{passive}}\|\mathbf{h}_{\mathrm{RI}}\|_{2}^{2}\|\mathbf{h}_{\mathrm{IT}}\|_{2}^{2}}{\sigma_{\mathrm{R}}^{2}}\overset{N_{\mathrm{I}}\nearrow}{\approx}\beta\frac{\Gamma^{4}(N_{\mathrm{G}}+\frac{1}{2})N_{\mathrm{I}}^{2}}{N_{\mathrm{G}}^{2}\Gamma^{4}(N_{\mathrm{G}})},\label{eq:asymp_snr_BD_passive_gen}
\end{equation}
and that for an extreme case of $G=1$ is 
\begin{equation}
\tilde{\bar{\gamma}}^{\mathsf{passive,BD}}=\frac{P_{\mathrm{T}}^{\mathsf{passive}}\|\mathbf{h}_{\mathrm{RI}}\|_{2}^{2}\|\mathbf{h}_{\mathrm{IT}}\|_{2}^{2}}{\sigma_{\mathrm{R}}^{2}}\overset{N_{\mathrm{I}}\nearrow}{\approx}\beta N_{\mathrm{I}}^{2}.\label{eq:asymp_snr_BD_passive}
\end{equation}
Comparing (\ref{eq:asymp_snr_BD_gen}), (\ref{eq:asymp_snr_BD}) and
(\ref{eq:asymp_snr_D})-(\ref{eq:asymp_snr_BD_passive}), we observe
that, with the same transmit power $P_{\mathrm{T}}$, the maximum
SNR gain of active BD-RIS over active D-RIS when $N_{\mathrm{I}}\rightarrow\infty$
is $\frac{16}{\pi^{2}}$, which is consistent with the maximum gain
of passive BD-RIS over passive D-RIS. \end{remark}

\begin{remark} From (\ref{eq:asymp_snr_BD}), (\ref{eq:asymp_snr_D}),
and (\ref{eq:asymp_snr_D_passive}), (\ref{eq:asymp_snr_BD_passive})
we observe that the asymptotic SNR for passive RISs grows quadratically
with $N_{\mathrm{I}}$ while the SNR for active RISs grows linearly
with $N_{\mathrm{I}}$. This is because adding power amplifiers in
RIS also introduces noise, of which the total power also grows linearly
with the number of elements. Such trend indicates that when $N_{\mathrm{I}}$
is sufficiently large, passive RISs will outperform active ones. The
required $N_{\mathrm{I}}$ for an active D-RIS to outperform a passive
D-RIS, i.e. $\bar{\gamma}^{\mathsf{active,D}}\ge\bar{\gamma}^{\mathsf{passive,D}}$,
has been derived in \cite{Zhang2023} as 
\begin{equation}
\begin{aligned}\bar{N}_{\mathrm{I}} & \le\frac{P_{\mathrm{T}}}{P_{\mathrm{T}}^{\mathsf{passive}}}\frac{P_{\mathrm{A}}\sigma_{\mathrm{R}}^{2}}{\sigma_{\mathrm{I}}^{2}P_{\mathrm{A}}\frac{1}{N_{\mathrm{I}}}\|\mathbf{h}_{\mathrm{RI}}\|_{2}^{2}+\sigma_{\mathrm{R}}^{2}P_{\mathrm{T}}\frac{1}{N_{\mathrm{I}}}\|\mathbf{h}_{\mathrm{IT}}\|_{2}^{2}+\sigma_{\mathrm{R}}^{2}\sigma_{\mathrm{I}}^{2}}\\
 & \overset{N_{\mathrm{I}}\nearrow}{\approx}\frac{P_{\mathrm{T}}}{P_{\mathrm{T}}^{\mathsf{passive}}}\frac{\sigma_{\mathrm{R}}^{2}P_{\mathrm{A}}}{\sigma_{\mathrm{I}}^{2}P_{\mathrm{A}}\zeta_{\mathrm{RI}}^{2}+\sigma_{\mathrm{R}}^{2}P_{\mathrm{T}}\zeta_{\mathrm{IT}}^{2}+\sigma_{\mathrm{R}}^{2}\sigma_{\mathrm{I}}^{2}},
\end{aligned}
\end{equation}
with $\frac{1}{N_{\mathrm{I}}}\|\mathbf{h}_{\mathrm{IT}}\|_{2}^{2}\overset{N_{\mathrm{I}}\nearrow}{\approx}\zeta_{\mathrm{RI}}^{2}$
and $\frac{1}{N_{\mathrm{I}}}\|\mathbf{h}_{\mathrm{IT}}\|_{2}^{2}\overset{N_{\mathrm{I}}\nearrow}{\approx}\zeta_{\mathrm{IT}}^{2}$.
This, according to Remark 1, also holds for $\tilde{\bar{\gamma}}^{\mathsf{active,BD}}\ge\tilde{\bar{\gamma}}^{\mathsf{passive,BD}}$.
Meanwhile, the required $N_{\mathrm{I}}$ for an active BD-RIS to
outperform passive D-RIS is given by 
\begin{equation}
\begin{aligned}\tilde{N}_{\mathrm{I}} & \le\frac{P_{\mathrm{T}}}{P_{\mathrm{T}}^{\mathsf{passive}}}\frac{P_{\mathrm{A}}\sigma_{\mathrm{R}}^{2}}{\sigma_{\mathrm{I}}^{2}P_{\mathrm{A}}\frac{1}{N_{\mathrm{I}}}\|\mathbf{h}_{\mathrm{RI}}\|_{2}^{2}+\sigma_{\mathrm{R}}^{2}P_{\mathrm{T}}\frac{1}{N_{\mathrm{I}}}\|\mathbf{h}_{\mathrm{IT}}\|_{2}^{2}+\sigma_{\mathrm{R}}^{2}\sigma_{\mathrm{I}}^{2}}\\
 & ~~~~\times\frac{\|\mathbf{h}_{\mathrm{RI}}\|_{2}^{2}\|\mathbf{h}_{\mathrm{IT}}\|_{2}^{2}}{\Big(\sum_{n=1}^{N_{\mathrm{I}}}|[\mathbf{h}_{\mathrm{RI}}]_{n}||[\mathbf{h}_{\mathrm{IT}}]_{n}|\Big)^{2}}\\
 & \overset{N_{\mathrm{I}}\nearrow}{\approx}\frac{P_{\mathrm{T}}}{P_{\mathrm{T}}^{\mathsf{passive}}}\frac{16\sigma_{\mathrm{R}}^{2}P_{\mathrm{A}}}{\pi^{2}(\sigma_{\mathrm{I}}^{2}P_{\mathrm{A}}\zeta_{\mathrm{RI}}^{2}+\sigma_{\mathrm{R}}^{2}P_{\mathrm{T}}\zeta_{\mathrm{IT}}^{2}+\sigma_{\mathrm{R}}^{2}\sigma_{\mathrm{I}}^{2})}.
\end{aligned}
\end{equation}
Therefore, we have $\tilde{N}_{\mathrm{I}}\overset{N_{\mathrm{I}}\nearrow}{\approx}\frac{16}{\pi^{2}}\bar{N}_{\mathrm{I}}$,
indicating that the active BD-RIS maintains its benefit over passive
D-RIS for a larger range of possible $N_{\mathrm{I}}$. \end{remark}

\begin{figure}
\centering \includegraphics[width=0.48\textwidth]{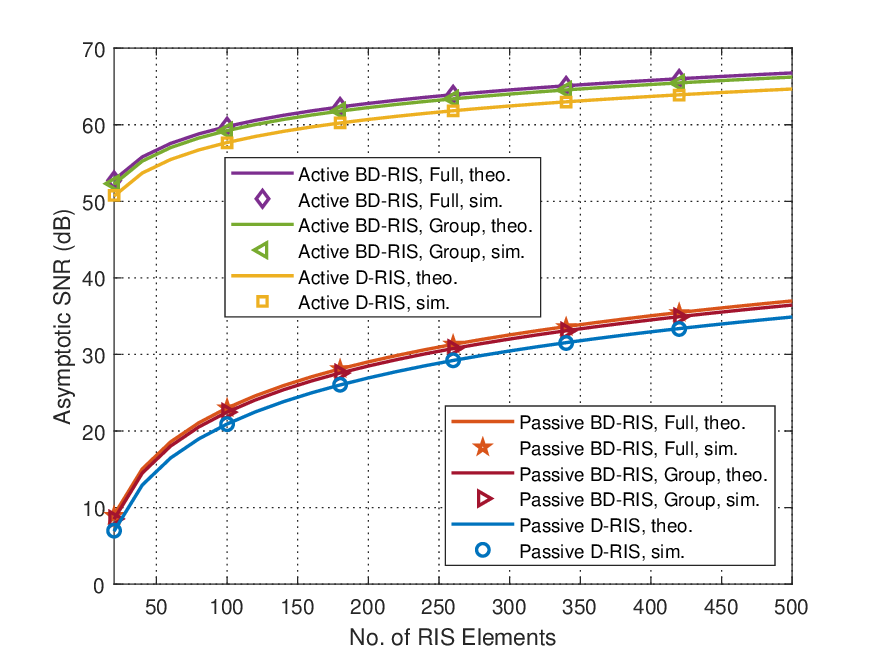}
\caption{SNR versus the number of elements $N_{\mathrm{I}}$. For the group-connected
active/passive BD-RIS, the group size is fixed to $N_{\mathrm{G}}=4$.}
\label{fig:snr_M}
\end{figure}

\begin{figure}
\centering \includegraphics[width=0.48\textwidth]{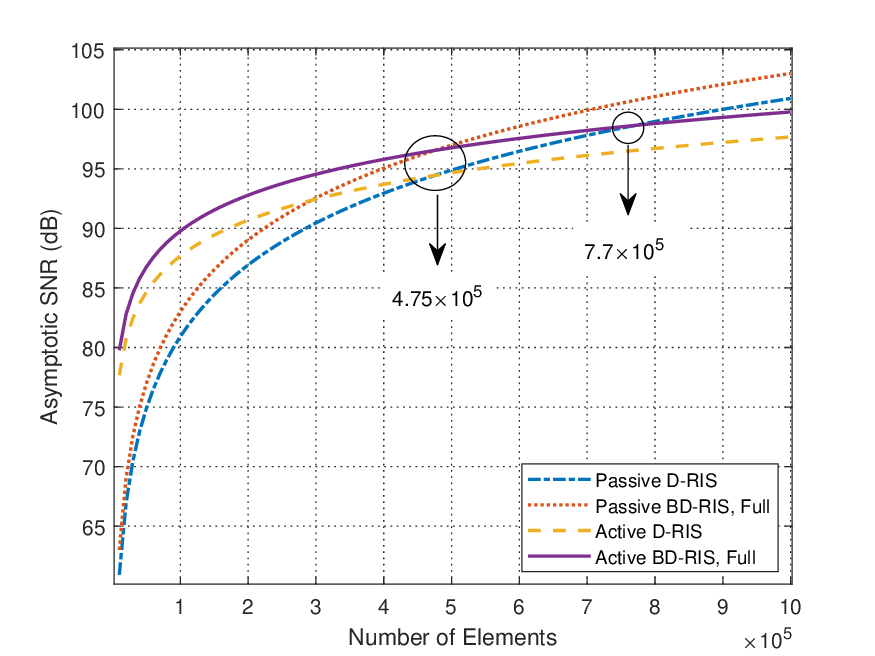}
\caption{Asymptotic SNR versus the number of elements $N_{\mathrm{I}}$.}
\label{fig:snr_M_asymp}
\end{figure}

To visualize the above results, we consider a case as follows. The
transmit power for passive BD-RIS aided systems is set as $P_{\mathrm{T}}^{\mathsf{passive}}=2$
W, while the transmit power and reflected power for active BD-RIS
aided systems are respectively given as $P_{\mathrm{T}}=1.9$ W and
$P_{\mathrm{A}}=0.1$ W. The noise power and path loss are respectively
set as $\sigma_{\mathrm{R}}^{2}=\sigma_{\mathrm{I}}^{2}=-90$ dBm
and $\zeta_{\mathrm{RI}}^{2}=\zeta_{\mathrm{IT}}^{2}=-70$ dB. Fig.
\ref{fig:snr_M} plots the SNR versus the number of elements based
on the asymptotic results in (\ref{eq:asymp_snr_BD_gen}), (\ref{eq:asymp_snr_BD}),
(\ref{eq:asymp_snr_D})-(\ref{eq:asymp_snr_BD_passive}), and the
numerical results using the closed-form solutions for D-RIS, i.e.
(\ref{eq:A}), (\ref{eq:D_RIS}), and BD-RIS, i.e. (\ref{eq:A}),
(\ref{eq:BD_RIS}). Results show that the asymptotic results (marked
as theo.) align well with numerical simulations (marked as sim.) and
that the maximum SNR gain of active BD-RIS over active D-RIS is the
same as that of passive BD-RIS over passive D-RIS (around 2 dB), consistent
with Remark 1. 
In addition, we visualize the crossing points after which passive RISs will outperform active RISs in Fig. \ref{fig:snr_M_asymp}. We can observe from Fig. \ref{fig:snr_M_asymp} that
the required numbers of elements for $\bar{\gamma}^{\mathsf{active,D}}\ge\bar{\gamma}^{\mathsf{passive,D}}$
($\tilde{\bar{\gamma}}^{\mathsf{active,BD}}\ge\tilde{\bar{\gamma}}^{\mathsf{passive,BD}}$)
and $\tilde{\bar{\gamma}}^{\mathsf{active,BD}}\ge\bar{\gamma}^{\mathsf{passive,D}}$
are respectively given as $\bar{N}_{\mathrm{I}}\lessapprox4.75\times10^{5}$
and $\tilde{N}_{\mathrm{I}}\lessapprox7.7\times10^{5}$, which are consistent with the theoretical results in Remark 2. Note
that such a huge number of elements is highly impractical based on
current technology. Therefore, active BD-RIS (D-RIS) has clear benefits
over passive BD-RIS (D-RIS).

\section{Active BD-RIS Aided MIMO System}

We next consider the active BD-RIS aided MIMO system with the model
(\ref{eq:channel2}). Assuming a linear digital precoder $\mathbf{F}\in\mathbb{C}^{N_{\mathrm{T}}\times N_{\mathrm{S}}}$
is adopted at the transmitter, where $N_{\mathrm{S}}$ denotes the
number of data streams, $\mathbf{x}$ can be rewritten as $\mathbf{x}=\mathbf{F}\mathbf{s}$.
Specifically, $\mathbf{s}\in\mathbb{C}^{N_{\mathrm{S}}\times1}$ denotes
the transmit symbol vector with $\mathbb{E}\{\mathbf{s}\mathbf{s}^{\mathsf{H}}\}=\mathbf{I}_{N_{\mathrm{S}}}$.
Therefore, the precoder $\mathbf{F}$ is constrained by $\|\mathbf{F}\|_{\mathsf{F}}^{2}\le P_{\mathrm{T}}$.
According to (\ref{eq:channel2}), the received signal is rewritten
as 
\begin{equation}\label{eq:received_signal}
\mathbf{y}=(\mathbf{H}_{\mathrm{RT}}+\mathbf{H}_{\mathrm{RI}}\mathbf{\Theta}\mathbf{H}_{\mathrm{IT}})\mathbf{F}\mathbf{s}+\mathbf{H}_{\mathrm{RI}}\mathbf{\Theta}\mathbf{n}_{\mathrm{I}}+\mathbf{n}_{\mathrm{R}}.
\end{equation}
Assuming perfect CSI at the transmitter\footnote{From (\ref{eq:received_signal}) we observe that separate information of $\mathbf{H}_\mathrm{RT}$, $\mathbf{H}_\mathrm{RI}$, and $\mathbf{H}_\mathrm{IT}$ is required for beamforming optimization. Note that active BD-RISs do not have signal processing capabilities but only scatter and amplify the incident signals to strengthen the wireless channel. In this case, installing a small number of low-cost sensing devices to active BD-RIS can help to separately acquire $\mathbf{H}_\mathrm{RI}$ and $\mathbf{H}_\mathrm{IT}$.}, the spectral efficiency is
given by 
\begin{equation}
R=\log_{2}(\det(\mathbf{I}_{N_{\mathrm{R}}}+\mathbf{R}^{-1}\mathbf{H}\mathbf{F}\mathbf{F}^{\mathsf{H}}\mathbf{H}^{\mathsf{H}})),
\end{equation}
where $\mathbf{H}=\mathbf{H}_{\mathrm{RT}}+\mathbf{H}_{\mathrm{RI}}\mathbf{\Theta}\mathbf{H}_{\mathrm{IT}}$
and $\mathbf{R}=\sigma_{\mathrm{I}}^{2}\mathbf{H}_{\mathrm{RI}}\mathbf{\Theta}\mathbf{\Theta}^{\mathsf{H}}\mathbf{H}_{\mathrm{RI}}^{\mathsf{H}}+\sigma_{\mathrm{R}}^{2}\mathbf{I}_{N_{\mathrm{R}}}$.
Meanwhile, the radiated power of active BD-RIS is constrained by 
\begin{equation}
\begin{aligned} & \mathbb{E}\{\|\mathbf{\Theta}\mathbf{H}_{\mathrm{IT}}\mathbf{F}\mathbf{s}+\mathbf{\Theta}\mathbf{n}_{\mathrm{I}}\|_{2}^{2}\}\\
 & ~~~~=\|\mathbf{\Theta}\mathbf{H}_{\mathrm{IT}}\mathbf{F}\|_{\mathsf{F}}^{2}+\sigma_{\mathrm{I}}^{2}\|\mathbf{\Theta}\|_{\mathsf{F}}^{2}\le P_{\mathrm{A}}.
\end{aligned}
\end{equation}
Then, the spectral efficiency maximization problem can be formulated
as 
\begin{subequations}
\label{eq:p0_mimo} 
\begin{align}
\max_{\mathbf{\Theta},\mathbf{F}}~ & R\\
\textrm{s.t.}~ & \|\mathbf{\Theta}\mathbf{H}_{\mathrm{IT}}\mathbf{F}\|_{\mathsf{F}}^{2}+\sigma_{\mathrm{I}}^{2}\|\mathbf{\Theta}\|_{\mathsf{F}}^{2}\le P_{\mathrm{A}},\label{eq:constraint_theta_mimo}\\
 & \text{(\ref{eq:constraint_theta2})},\\
 & \|\mathbf{F}\|_{\mathsf{F}}^{2}\le P_{\mathrm{T}},\label{eq:constraint_power_mimo}
\end{align}
\end{subequations}
 which is a typical multi-variable optimization with complex objective
functions and constraints. Below, we propose an iterative solution
with theoretically guaranteed convergence.

\subsection{Iterative Algorithm}

\label{subsec:mimo} The main difficulty of solving problem (\ref{eq:p0_mimo})
lies in the complicated objective function $R$ that includes matrix
inverse and determinant operation and the constraint (\ref{eq:constraint_theta_mimo})
coupling $\mathbf{\Theta}$ and $\mathbf{F}$. To simplify the design,
we first adopt the well-known weighted minimum mean square error (WMMSE)-rate
relationship \cite{christensen2008weighted}. Let $\widehat{\mathbf{s}}=\mathbf{W}^{\mathsf{H}}\mathbf{y}$
be the estimate of $\mathbf{s}$ with a combiner $\mathbf{W}\in\mathbb{C}^{N_{\mathrm{R}}\times N_{\mathrm{S}}}$
at the receiver. Define the MSE matrix as 
\begin{equation}
\begin{aligned}\mathbf{E} & =\mathbb{E}\{(\widehat{\mathbf{s}}-\mathbf{s})(\widehat{\mathbf{s}}-\mathbf{s})^{\mathsf{H}}\}\\
 & =\mathbf{W}^{\mathsf{H}}\mathbf{H}\mathbf{F}\mathbf{F}^{\mathsf{H}}\mathbf{H}^{\mathsf{H}}\mathbf{W}-2\Re\{\mathbf{W}^{\mathsf{H}}\mathbf{H}\mathbf{F}\}+\mathbf{W}^{\mathsf{H}}\mathbf{R}\mathbf{W}+\mathbf{I}_{N_{\mathrm{S}}}.
\end{aligned}
\end{equation}
Then the optimal combiner that achieves the minimum MSE, $\min_{\mathbf{W}}\mathsf{Tr}(\mathbf{E})$,
is 
\begin{equation}
\mathbf{W}^{\star}=(\mathbf{H}\mathbf{F}\mathbf{F}^{\mathsf{H}}\mathbf{H}^{\mathsf{H}}+\mathbf{R})^{-1}\mathbf{H}\mathbf{F},\label{eq:w}
\end{equation}
yielding $\mathbf{E}^{\star}=\mathbf{I}_{N_{\mathrm{S}}}-\mathbf{F}^{\mathsf{H}}\mathbf{H}^{\mathsf{H}}\mathbf{W}^{\star}=(\mathbf{I}_{N_{\mathrm{S}}}+\mathbf{F}^{\mathsf{H}}\mathbf{H}^{\mathsf{H}}\mathbf{R}^{-1}\mathbf{H}\mathbf{F})^{-1}$.
We further introduce a weighting matrix $\mathbf{U}\in\mathbb{C}^{N_{\mathrm{S}}\times N_{\mathrm{S}}}$.
Then we can establish the WMMSE-rate relationship as $R=\max_{\mathbf{U}}\log_{2}(\det(\mathbf{U}))-\mathsf{Tr}(\mathbf{U}\mathbf{E}^{\star})+N_{\mathrm{S}}$,
where the maximum is achieved when 
\begin{equation}
\mathbf{U}^{\star}=(\mathbf{E}^{\star})^{-1}=\mathbf{I}_{N_{\mathrm{S}}}+\mathbf{F}^{\mathsf{H}}\mathbf{H}^{\mathsf{H}}\mathbf{R}^{-1}\mathbf{H}\mathbf{F}.\label{eq:u}
\end{equation}
This motivates to reformulate problem (\ref{eq:p0_mimo}) as 
\begin{subequations}
\label{eq:p1_mimo} 
\begin{align}
\max_{\mathbf{U},\mathbf{W},\mathbf{\Theta},\mathbf{F}} & \log_{2}(\det(\mathbf{U}))-\mathsf{Tr}(\mathbf{U}\mathbf{E})\\
\textrm{s.t.}~~ & \text{(\ref{eq:constraint_theta_mimo}),~(\ref{eq:constraint_theta2}),~(\ref{eq:constraint_power_mimo})}.
\end{align}
\end{subequations}
 Problem (\ref{eq:p1_mimo}) will be solved in an iterative manner
by updating each variable while fixing others. The solutions to $\mathbf{W}$
and $\mathbf{U}$ have been given respectively in (\ref{eq:w}) and
(\ref{eq:u}). Solutions to $\mathbf{\Theta}$ and $\mathbf{F}$ will
be elaborated below.

\subsubsection{Update $\mathbf{\Theta}$}
With fixed $\mathbf{U}$, $\mathbf{W}$, and $\mathbf{F}$, the $\mathbf{\Theta}$-subproblem can be derived from (\ref{eq:p1_mimo}) as 
 \begin{equation}\label{eq:p1_theta}
    \begin{aligned}
        \min_{\mathbf{\Theta}}~&\mathsf{Tr}(\mathbf{\Theta}\mathbf{H}_\mathrm{IT}\mathbf{F}\mathbf{F}^\mathsf{H}\mathbf{H}_\mathrm{IT}^\mathsf{H}\mathbf{\Theta}^\mathsf{H}\mathbf{H}_\mathrm{RI}^\mathsf{H}\mathbf{W}\mathbf{U}\mathbf{W}^\mathsf{H}\mathbf{H}_\mathrm{RI}\\
        &~~~ + \sigma_\mathrm{I}^2\mathbf{\Theta}^\mathsf{H}\mathbf{H}_\mathrm{RI}^\mathsf{H}\mathbf{W}\mathbf{U}\mathbf{W}^\mathsf{H}\mathbf{H}_\mathrm{RI}\mathbf{\Theta}\\
        &~~~- 2\Re\{\mathbf{\Theta}^\mathsf{H}(\mathbf{H}_\mathrm{RI}^\mathsf{H}\mathbf{W}\mathbf{U}(\mathbf{I}_{N_\mathrm{S}} - \mathbf{W}^\mathsf{H}\mathbf{H}_\mathrm{RT}\mathbf{F})\mathbf{F}^\mathsf{H}\mathbf{H}_\mathrm{IT}^\mathsf{H})\})\\
        \mathrm{s.t.}~ &\mathsf{Tr}(\mathbf{\Theta}\mathbf{H}_\mathrm{IT}\mathbf{F}\mathbf{F}^\mathsf{H}\mathbf{H}_\mathrm{IT}^\mathsf{H}\mathbf{\Theta}^\mathsf{H} + \sigma_\mathrm{I}^2\mathbf{\Theta}^\mathsf{H}\mathbf{\Theta}) \le P_\mathrm{A},\\
        &\mathbf{\Theta} = \mathsf{blkdiag}(\mathbf{\Theta}_1,\ldots,\mathbf{\Theta}_G), \mathbf{\Theta}_g\in\mathcal{T}_g,\forall g.
    \end{aligned}
\end{equation}
The constraint that $\mathbf{\Theta}$
is a block-diagonal matrix motivates us to extract terms related to
only the diagonal blocks, i.e. $\mathbf{\Theta}_{1},\ldots,\mathbf{\Theta}_{G}$, in $\mathbf{\Theta}$. Therefore, we block $\mathbf{H}_\mathrm{RI}$ and $\mathbf{H}_\mathrm{IT}$ respectively as $\mathbf{H}_\mathrm{RI} = [\mathbf{H}_{\mathrm{RI},1},\ldots,\mathbf{H}_{\mathrm{RI},G}]$ and $\mathbf{H}_\mathrm{IT} = [\mathbf{H}_{\mathrm{IT},1}^\mathsf{T},\ldots,\mathbf{H}_{\mathrm{IT},G}^\mathsf{T}]^\mathsf{T}$ with $\mathbf{H}_{\mathrm{RI},g}=[\mathbf{H}_{\mathrm{RI}}]_{:,(g-1)N_{\mathrm{G}}+1:gN_{\mathrm{G}}}$
and $\mathbf{H}_{\mathrm{IT},g}=[\mathbf{H}_{\mathrm{IT}}]_{(g-1)N_{\mathrm{G}}+1:gN_{\mathrm{G}},:}$, $\forall g$, and reformulate (\ref{eq:p1_theta}) as
 \begin{equation}
    \begin{aligned}
        \min_{\mathbf{\Theta}}~&\mathsf{Tr}\Big(\sum_{g=1}^G\sum_{g'=1}^G\mathbf{\Theta}_{g'}\mathbf{H}_{\mathrm{IT},g'}\mathbf{F}\mathbf{F}^\mathsf{H}\mathbf{H}_{\mathrm{IT},g}^\mathsf{H}\mathbf{\Theta}_{g}^\mathsf{H}\mathbf{H}_{\mathrm{RI},g}^\mathsf{H}\mathbf{W}\\
        &~~~\times\mathbf{U}\mathbf{W}^\mathsf{H}\mathbf{H}_{\mathrm{RI},g'}
         + \sigma_\mathrm{I}^2\sum_{g=1}^G\mathbf{\Theta}_g^\mathsf{H}\mathbf{H}_{\mathrm{RI},g}^\mathsf{H}\mathbf{W}\mathbf{U}\mathbf{W}^\mathsf{H}\mathbf{H}_{\mathrm{RI},g}\mathbf{\Theta}_g\\
        &~~~- 2\sum_{g=1}^G\Re\{\mathbf{\Theta}_g^\mathsf{H}(\mathbf{H}_{\mathrm{RI},g}^\mathsf{H}\mathbf{W}\mathbf{U}(\mathbf{I}_{N_\mathrm{S}} - \mathbf{W}^\mathsf{H}\mathbf{H}_\mathrm{RT}\mathbf{F})\\
        &~~~\times\mathbf{F}^\mathsf{H}\mathbf{H}_{\mathrm{IT},G}^\mathsf{H})\}\Big)\\
        \mathrm{s.t.}~ &\mathsf{Tr}\Big(\sum_{g=1}^G\big(\mathbf{\Theta}_g\mathbf{H}_{\mathrm{IT},g}\mathbf{F}\mathbf{F}^\mathsf{H}\mathbf{H}_{\mathrm{IT},g}^\mathsf{H}\mathbf{\Theta}_g^\mathsf{H} + \sigma_\mathrm{I}^2\mathbf{\Theta}_g^\mathsf{H}\mathbf{\Theta}_g\big)\Big) \le P_\mathrm{A},\\
        &\mathbf{\Theta}_g\in\mathcal{T}_g,\forall g.
    \end{aligned}\label{eq:p2_theta}
\end{equation}

For non-reciprocal active BD-RIS where each $\mathbf{\Theta}_g$ can be any complex matrices, we can reformulate (\ref{eq:p2_theta}) as 
\begin{equation}
\begin{aligned}\min_{\bm{\theta}}~ & \bm{\theta}^{\mathsf{H}}(\mathbf{B}_{1}+\mathbf{B}_{2})\bm{\theta}-2\Re\{\bm{\theta}^{\mathsf{H}}\mathbf{c}\}~\textrm{s.t.}~\bm{\theta}^{\mathsf{H}}\mathbf{D}\bm{\theta}\le P_{\mathrm{A}},\end{aligned}
\label{eq:subprob_theta_nr_mimo}
\end{equation}
where 
\begin{equation}
\begin{aligned}\bm{\theta} & =[\bm{\theta}_{1}^{\mathsf{T}},\ldots,\bm{\theta}_{G}^{\mathsf{T}}]^{\mathsf{T}},~\bm{\theta}_{g}=\mathsf{vec}(\mathbf{\Theta}_{g}),\\
\mathbf{B}_{1} & =\left[\begin{matrix}\mathbf{B}_{1,1,1} & \cdots & \mathbf{B}_{1,1,G}\\
\vdots & \ddots & \vdots\\
\mathbf{B}_{1,G,1} & \cdots & \mathbf{B}_{1,G,G}
\end{matrix}\right],\\
\mathbf{B}_{1,g,g'} & =(\mathbf{H}_{\mathrm{IT},g'}\mathbf{F}\mathbf{F}^{\mathsf{H}}\mathbf{H}_{\mathrm{IT},g}^{\mathsf{H}})^{\mathsf{T}}\otimes(\mathbf{H}_{\mathrm{RI},g}^{\mathsf{H}}\mathbf{W}\mathbf{U}\mathbf{W}^{\mathsf{H}}\mathbf{H}_{\mathrm{RI},g'}),\\
\mathbf{B}_{2} & =\mathsf{blkdiag}(\mathbf{B}_{2,1},\ldots,\mathbf{B}_{2,G}),\\
\mathbf{B}_{2,g} & =\sigma_{\mathrm{I}}^{2}\mathbf{I}_{N_{\mathrm{G}}}\otimes(\mathbf{H}_{\mathrm{RI},g}^{\mathsf{H}}\mathbf{W}\mathbf{U}\mathbf{W}^{\mathsf{H}}\mathbf{H}_{\mathrm{RI},g}),\\
\mathbf{c} & =[\mathbf{c}_{1}^{\mathsf{T}},\ldots,\mathbf{c}_{G}^{\mathsf{T}}]^{\mathsf{T}},\\
\mathbf{c}_{g} & =\mathsf{vec}(\mathbf{H}_{\mathrm{RI},g}^{\mathsf{H}}\mathbf{W}\mathbf{U}(\mathbf{I}_{N_{\mathrm{S}}}-\mathbf{W}^{\mathsf{H}}\mathbf{H}_{\mathrm{RT}}\mathbf{F})\mathbf{F}^{\mathsf{H}}\mathbf{H}_{\mathrm{IT},g}^{\mathsf{H}}),\\
\mathbf{D} & =\mathsf{blkdiag}(\mathbf{D}_{1},\ldots,\mathbf{D}_{G}),\\
\mathbf{D}_{g} & =(\mathbf{H}_{\mathrm{IT},g}\mathbf{F}\mathbf{F}^{\mathsf{H}}\mathbf{H}_{\mathrm{IT},g}^{\mathsf{H}})^{\mathsf{T}}\otimes\mathbf{I}_{N_{\mathrm{G}}}+\sigma_{\mathrm{I}}^{2}\mathbf{I}_{N_{\mathrm{G}}^{2}}.
\end{aligned}
\end{equation}
Problem (\ref{eq:subprob_theta_nr_mimo}) is a standard
QCQP and can be solved by applying the Lagrangian multiplier method
with a closed-form solution 
\begin{equation}
\bm{\theta}^{\star}=(\mathbf{B}_{1}+\mathbf{B}_{2}+\mu^{\star}\mathbf{D})^{-1}\mathbf{c},\label{eq:theta_nr}
\end{equation}
where $\mu^{\star}\ge0$ can be obtained by bisection search. This
gives the solution $\mathbf{\Theta}^{\star}=\mathsf{blkdiag}(\mathsf{unvec}(\bm{\theta}_{1}^{\star}),\ldots,\mathsf{unvec}(\bm{\theta}_{G}^{\star}))$.

For reciprocal active BD-RIS, the symmetric property of each $\mathbf{\Theta}_{g}$
suggests that $\mathbf{\Theta}_{g}$ is essentially determined by
its diagonal and lower-triangular (or upper-triangular) entries. Therefore,
we have $\bm{\theta}_{g}=\mathbf{P}\bar{\bm{\theta}}_{g}$, where
$\bar{\bm{\theta}}_{g}\in\mathbb{C}^{\frac{N_{\mathrm{G}}(N_{\mathrm{G}}+1)}{2}\times1}$
contains the diagonal and lower-triangular entries of $\mathbf{\Theta}_{g}$
and $\mathbf{P}\in\{0,1\}^{N_{\mathrm{G}}^{2}\times\frac{N_{\mathrm{G}}(N_{\mathrm{G}}+1)}{2}}$
is a binary matrix mapping $\bar{\bm{\theta}}_{g}$ into $\bm{\theta}_{g}$.
The binary matrix $\mathbf{P}$ is defined as 
\begin{equation}
[\mathbf{P}]_{N_{\mathrm{G}}(i-1)+i',l}=\begin{cases}
1, & l=\frac{i(i-1)}{2}+i'\text{~and~}1\le i'\le i,\\
1, & l=\frac{i'(i'-1)}{2}+i\text{~and~}i<i'\le N_{\mathrm{G}},\\
0, & \text{otherwise}.
\end{cases}
\end{equation}
It turns out that it is sufficient to solve the $\bar{\bm{\theta}}$-subproblem
\begin{equation}
\begin{aligned}\min_{\bar{\bm{\theta}}}~ & \bar{\bm{\theta}}^{\mathsf{H}}(\bar{\mathbf{B}}_{1}+\bar{\mathbf{B}}_{2})\bar{\bm{\theta}}-2\Re\{\bar{\bm{\theta}}^{\mathsf{H}}\bar{\mathbf{c}}\}~\mathrm{s.t.}~\bar{\bm{\theta}}^{\mathsf{H}}\bar{\mathbf{D}}\bar{\bm{\theta}}\le P_{A},\end{aligned}
\label{eq:subprob_theta_r_mimo}
\end{equation}
where 
\begin{equation}
\begin{aligned}\bar{\bm{\theta}} & =[\bar{\bm{\theta}}_{1}^{\mathsf{T}},\ldots,\bar{\bm{\theta}}_{G}^{\mathsf{T}}]^{\mathsf{T}},\\
\bar{\mathbf{B}}_{1} & =\left[\begin{matrix}\bar{\mathbf{B}}_{1,1,1} & \ldots & \bar{\mathbf{B}}_{1,1,G}\\
\vdots & \ddots & \vdots\\
\bar{\mathbf{B}}_{1,G,1} & \ldots & \bar{\mathbf{B}}_{1,G,G}
\end{matrix}\right],\bar{\mathbf{B}}_{1,g,g'}=\mathbf{P}^{\mathsf{T}}\mathbf{B}_{1,g,g'}\mathbf{P},\\
\bar{\mathbf{B}}_{2} & =\mathsf{blkdiag}(\bar{\mathbf{B}}_{2,1},\ldots,\bar{\mathbf{B}}_{2,G}),\bar{\mathbf{B}}_{2,g,}=\mathbf{P}^{\mathsf{T}}\mathbf{B}_{2,g,}\mathbf{P},\\
\bar{\mathbf{c}} & =[\bar{\mathbf{c}}_{1}^{\mathsf{T}},\ldots,\bar{\mathbf{c}}_{G}^{\mathsf{T}}]^{\mathsf{T}},\bar{\mathbf{c}}_{g}=\mathbf{P}^{\mathsf{T}}\mathbf{c}_{g},\\
\bar{\mathbf{D}} & =\mathsf{blkdiag}(\bar{\mathbf{D}}_{1},\ldots,\bar{\mathbf{D}}_{G}),\bar{\mathbf{D}}_{g}=\mathbf{P}^{\mathsf{T}}\mathbf{D}_{g}\mathbf{P}.
\end{aligned}
\end{equation}
Problem (\ref{eq:subprob_theta_nr_mimo}) is again a standard QCQP,
whose closed-form solution can be found by applying the Lagrangian
multiplier method: 
\begin{equation}
\bar{\bm{\theta}}^{\star}=(\bar{\mathbf{B}}_{1}+\bar{\mathbf{B}}_{2}+\bar{\mu}^{\star}\bar{\mathbf{D}})^{-1}\bar{\mathbf{c}},\label{eq:theta_re}
\end{equation}
where $\bar{\mu}^{\star}\ge0$ can be obtained by a bisection search.
Therefore, we have $\mathbf{\Theta}^{\star}=\mathsf{blkdiag}(\mathsf{unvec}(\mathbf{P}\bar{\bm{\theta}}_{1}^{\star}),\ldots,\mathsf{unvec}(\mathbf{P}\bar{\bm{\theta}}_{G}^{\star}))$.

\subsubsection{Update $\mathbf{F}$}

The $\mathbf{F}$-subproblem is given by 
\begin{equation}
\begin{aligned}\min_{\mathbf{F}}~ & \mathsf{Tr}(\mathbf{F}^{\mathsf{H}}\mathbf{H}^{\mathsf{H}}\mathbf{W}\mathbf{U}\mathbf{W}^{\mathsf{H}}\mathbf{H}\mathbf{F}-2\Re\{\mathbf{F}^{\mathsf{H}}\mathbf{H}^{\mathsf{H}}\mathbf{W}\mathbf{U}\})\\
\mathrm{s.t.}~ & \mathsf{Tr}(\mathbf{F}^{\mathsf{H}}\mathbf{H}_{\mathrm{IT}}^{\mathsf{H}}\mathbf{\Theta}^{\mathsf{H}}\mathbf{\Theta}\mathbf{H}_{\mathrm{IT}}\mathbf{F})\le\bar{P}_{\mathrm{A}},\|\mathbf{F}\|_{\mathrm{F}}^{2}\le P_{\mathrm{T}},
\end{aligned}
\label{eq:subprob_precoder}
\end{equation}
where $\bar{P}_{\mathrm{A}}=P_{\mathrm{A}}-\sigma_{\mathrm{I}}^{2}\|\mathbf{\Theta}\|_{\mathrm{F}}^{2}$.
Problem (\ref{eq:subprob_precoder}) is also a QCQP and has a closed
form solution as 
\begin{equation}
\begin{aligned}\mathbf{F}^{\star}= & (\mathbf{H}^{\mathsf{H}}\mathbf{W}\mathbf{U}\mathbf{W}^{\mathsf{H}}\mathbf{H}+\mu_{1}^{\star}\mathbf{H}_{\mathrm{IT}}^{\mathsf{H}}\mathbf{\Theta}^{\mathsf{H}}\mathbf{\Theta}\mathbf{H}_{\mathrm{IT}}+\mu_{2}^{\star}\mathbf{I}_{N_{\mathrm{T}}})^{-1}\\
 & ~~~~~~\times\mathbf{H}^{\mathsf{H}}\mathbf{W}\mathbf{U},
\end{aligned}
\label{eq:f}
\end{equation}
where $\mu_{1}^{\star}\ge0$ and $\mu_{2}^{\star}\ge0$ can be obtained
by a two-dimensional grid search.

\subsection{Algorithm and Convergence}

With proper initialization of $\mathbf{\Theta}$ and $\mathbf{F}$,
the four variables $\mathbf{W}$, $\mathbf{U}$, $\mathbf{\Theta}$,
and $\mathbf{F}$ are updated, respectively based on (\ref{eq:w}),
(\ref{eq:u}), (\ref{eq:theta_nr}) for non-reciprocal architecture
or (\ref{eq:theta_re}) for reciprocal architecture, and (\ref{eq:f}),
in an iterative manner until convergence. The iteration will converge
to (at least) a local optimum due to the fact that the optimization
of each variable decreases the objective function, and that the objective
function of interest is bounded above.

\subsection{Computational Complexity Analysis}

The computational complexity for the proposed algorithm jointly depends
on the updates of the four blocks $\mathbf{W}$, $\mathbf{U}$, $\mathbf{\Theta}$,
and $\mathbf{P}$, whose computational complexity is all dominated
by the matrix inverse of different dimensions. Specifically, the complexity
of updating $\mathbf{W}$ based on (\ref{eq:w}) and $\mathbf{U}$
based on (\ref{eq:u}) are both $\mathcal{O}(N_{\mathrm{R}}^{3})$,
coming from calculating the inverse of $N_{\mathrm{R}}\times N_{\mathrm{R}}$
matrices. The complexity of updating $\mathbf{\Theta}$ for non-reciprocal
and reciprocal active BD-RIS are respectively given by $\mathcal{O}(I^{\mathsf{bi,NR}}N_{\mathrm{I}}^{3}N_{\mathrm{G}}^{3})$
and $\mathcal{O}(I^{\mathsf{bi,RE}}N_{\mathrm{I}}^{3}(\frac{N_{\mathrm{G}}+1}{2})^{3})$,
coming from calculating the inverse of $N_{\mathrm{I}}N_{\mathrm{G}}\times N_{\mathrm{I}}N_{\mathrm{G}}$
and $N_{\mathrm{I}}\frac{N_{\mathrm{G}}+1}{2}\times N_{\mathrm{I}}\frac{N_{\mathrm{G}}+1}{2}$
matrices, where $I^{\mathsf{bi,NR}}$ and $I^{\mathsf{bi,RE}}$ denote
the number of iterations for bisection search. The complexity of updating
$\mathbf{F}$ has a complexity of $\mathcal{O}(I^{\mathsf{grid}}N_{\mathrm{T}}^{3})$
due to the inverse of $N_{\mathrm{T}}\times N_{\mathrm{T}}$ matrices,
where $I^{\mathsf{grid}}$ denotes the number of iterations for grid
search. Thus, the overall computational complexity for the proposed
algorithm is given by $\mathcal{O}(I^{\mathsf{NR}}(2N_{\mathrm{R}}^{3}+I^{\mathsf{grid}}N_{\mathrm{T}}^{3}+I^{\mathsf{bi,NR}}N_{\mathrm{I}}^{3}N_{\mathrm{G}}^{3}))$
for non-reciprocal active BD-RIS and $\mathcal{O}(I^{\mathsf{RE}}(2N_{\mathrm{R}}^{3}+I^{\mathsf{grid}}N_{\mathrm{T}}^{3}+I^{\mathsf{bi,RE}}N_{\mathrm{I}}^{3}(\frac{N_{\mathrm{G}}+1}{2})^{3}))$
for reciprocal active BD-RIS, where $I^{\mathsf{NR}}$ and $I^{\mathsf{RE}}$
denote the number of iterations to achieve the convergence of the
proposed algorithm.

\section{Performance Evaluation}

To visualize the performance of active BD-RIS-aided wireless systems,
we present simulation results based on the following parameter settings.
The channels are modeled as the combination of small-scale fading
and large-scale fading. The small-scale fading is characterized by
the Rician fading with a Rician factor $\kappa=1$; the large-scale
fading is based on the 3GPP standard and is characterized by $\mathrm{PL}|_{\text{dB}}=41.2+28.7\log d$
\cite{Zhang2023}, where $d$ denotes the distance between devices.
This indicates that a strong direct link exists between the transmitter
and receiver. The transmitter, active BD-RIS, and receiver are respectively
located, based on a 2D coordinate, at {[}0, -60 m{]}, {[}300m, 10
m{]}, and {[}300 m, 0{]}. The noise powers are set as $\sigma_{\mathrm{R}}^{2}=\sigma_{\mathrm{I}}^{2}=-90$
dBm. For fair comparison, given a transmit power budget $P_{\mathrm{T}}^{\mathsf{tot}}$
for a wireless system without RIS, the transmit power and reflected
power for an active BD-RIS-aided system are respectively given by
$P_{\mathrm{T}}=0.99P_{\mathrm{T}}^{\mathsf{tot}}$ and $P_{\mathrm{A}}=0.01P_{\mathrm{T}}^{\mathsf{tot}}$.
Meanwhile, the transmit power for a passive BD-RIS-aided systems is
given by $P_{\mathrm{T}}^{\mathsf{tot}}$.

\begin{figure}
    \centering
    \subfigure[$N_\mathrm{T} = N_\mathrm{R} = N_\mathrm{S} = 2$, $N_\mathrm{I} = 32$]{
    \includegraphics[width=0.48\textwidth]{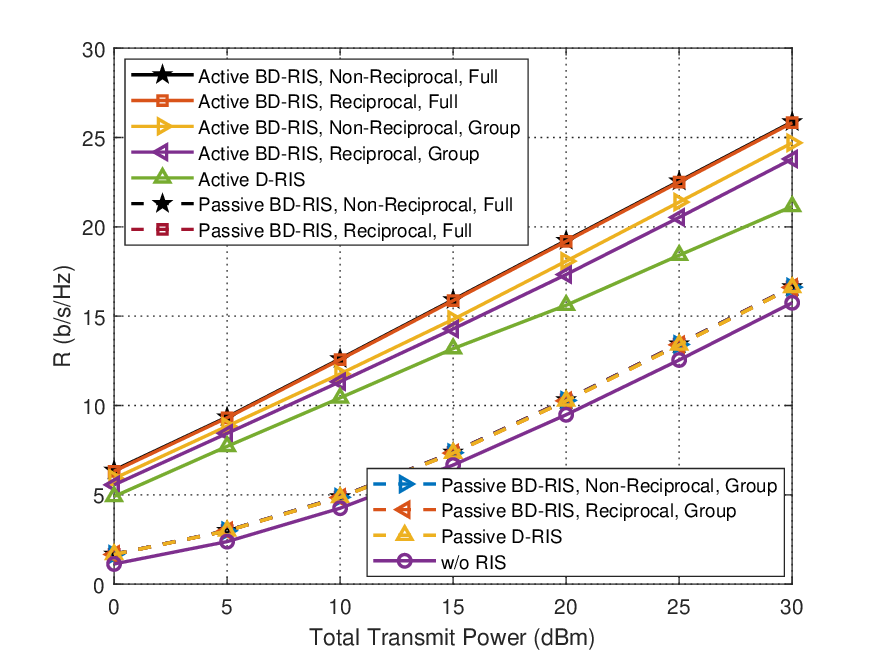}}
    \subfigure[$N_\mathrm{T} = N_\mathrm{R} = N_\mathrm{S} = 3$, $N_\mathrm{I} = 48$]{\includegraphics[width=0.48\textwidth]{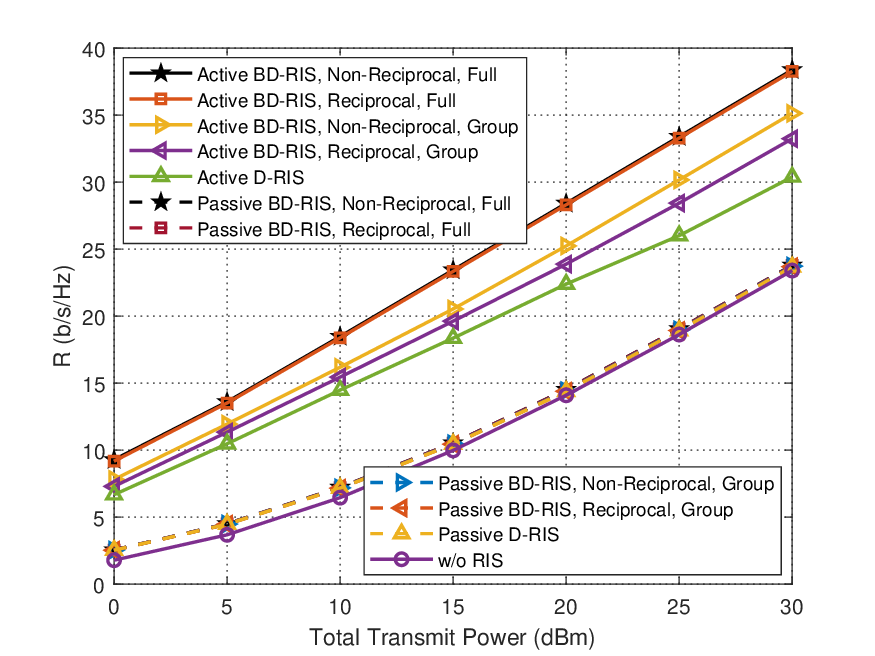}}
    \caption{Spectral efficiency versus transmit power $P_\mathrm{T}^\mathsf{tot}$. $N_\mathrm{G} = 2$ for group-connected architectures.}
    \label{fig:R_P_mimo}
\end{figure}

\begin{figure}
    \centering
    \subfigure[$N_\mathrm{T} = N_\mathrm{R} = N_\mathrm{S} = 2$]{
    \includegraphics[width=0.48\textwidth]{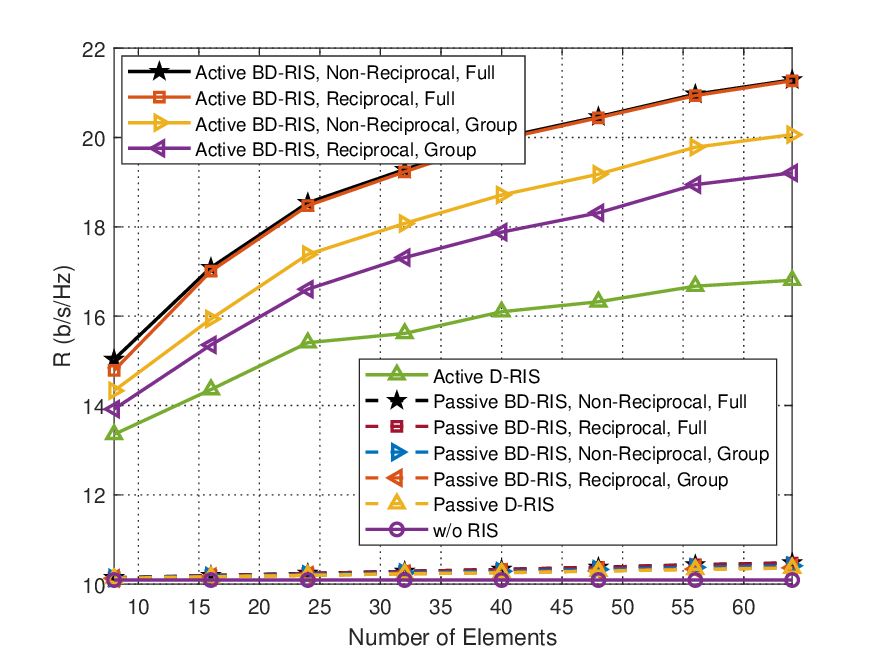}}
    \subfigure[$N_\mathrm{T} = N_\mathrm{R} = N_\mathrm{S} = 3$]{\includegraphics[width=0.48\textwidth]{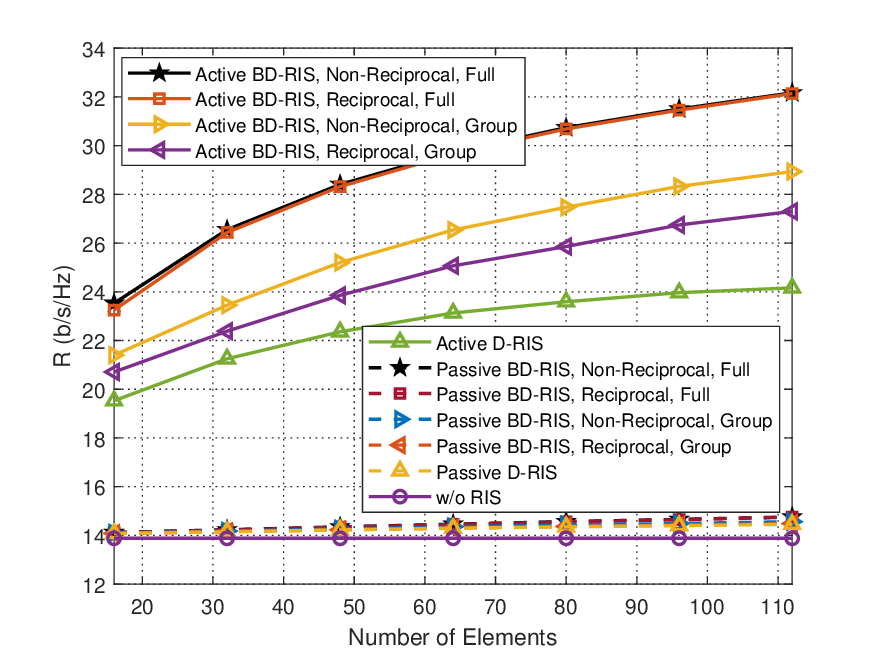}}
    \caption{Spectral efficiency versus the number of elements. $P_\mathrm{T}^\mathsf{tot} = 20$ dBm. $N_\mathrm{G} = 2$ for group-connected architectures.}
    \label{fig:R_M_mimo}
\end{figure}

Fig. \ref{fig:R_P_mimo} reports the spectral efficiency of an active
BD-RIS-aided MIMO system as a function of transmit power $P_{\mathrm{T}}^{\mathsf{tot}}$
for different parameter settings. For comparison purpose, we also
plot the performance for a passive BD-RIS-aided MIMO system. The optimization
for such a system can be done by reusing the WMMSE-rate relationship,
while the main difference lies in the optimization of BD-RIS scattering
matrices. For the case of passive BD-RIS with non-reciprocal architectures, $\mathbf{\Theta}$ is a complex unitary matrix. 
This allows us to use the well-known conjugate-gradient method applied on a manifold of complex unitary matrices \cite{li2022beyond}. For the case of
passive BD-RIS with reciprocal architectures, $\mathbf{\Theta}$ is complex symmetric and unitary, that is $\mathbf{\Theta} = \mathbf{\Theta}^\mathsf{T}$ and $\mathbf{\Theta}^\mathsf{H}\mathbf{\Theta} = \mathbf{I}_{N_\mathrm{I}}$. We adopt the Takagi factorization and decompose $\mathbf{\Theta}$ as $\mathbf{\Theta}=\mathbf{Q}\mathbf{Q}^{\mathsf{T}}$,
with $\mathbf{Q}$ being a complex unitary matrix \cite{santamaria2024mimo}. This allows us to skip the symmetric constraint of $\mathbf{\Theta}$ 
and directly optimize $\mathbf{Q}$ using the conjugate-gradient method
applied on a manifold of complex unitary matrices.
From Fig. \ref{fig:R_P_mimo} we
have the following observations. \textit{First}, introducing active
BD-RIS in MIMO systems can significantly improve the performance,
even when the direct channel from the transmitter and receiver exists
and is strong. However, in this case, applying passive BD-RIS in MIMO
systems can only provide marginal performance improvement. For example,
adding an 32-element active BD-RIS in a $2\times2$ MIMO system can
increase the spectral efficiency by around 10 b/s/Hz when the transmit
power is fixed to 30 dBm; adding an 32-element passive BD-RIS in the
same system can only increase the spectral efficiency by around 1
b/s/Hz. This result indicates that passive BD-RIS is more beneficial
in scenarios where the direct link is block, while active BD-RIS has
wider use cases. \textit{Second}, active BD-RIS-aided MIMO clearly
outperforms active D-RIS-aided MIMO, and the performance improvement
is more obvious for active BD-RIS with more complex architectures.
For example, an 48-element active BD-RIS with non-reciprocal fully-
and group-connected architectures can increase the spectral efficiency
for a $3\times3$ MIMO system by, respectively, around 27\% and 17\%
over an 48-element active D-RIS. This is because complex and diverse
channels are better captured by more BD-RIS architectures with more
inter-element interconnections. \textit{Third}, for active BD-RIS,
the performance improvement achieved by non-reciprocal architecture
compared to reciprocal architecture is not very prominent, especially
when the former requires much higher circuit complexity. For example,
reciprocal and non-reciprocal fully-connected architectures achieve
almost the same performance within the considered simulation range;
non-reciprocal group-connected architecture (with group size equal
to 2) can only slightly improve the spectral efficiency by around
7\% over reciprocal group-connected architecture for a $3\times3$
MIMO system. This result indicates that, under the same performance
requirement, using reciprocal active BD-RIS can be a more practical
choice in real-world situations.

Fig. \ref{fig:R_M_mimo} plots the spectral efficiency of an active
BD-RIS-aided MIMO system versus the number of elements $N_{\mathrm{I}}$.
Similar conclusions can be obtained from Fig. \ref{fig:R_M_mimo}
that active BD-RIS always outperforms active D-RIS and passive BD-RIS.
For example, for a $2\times2$ MIMO system, replacing a passive BD-RIS
with an active BD-RIS having non-reciprocal fully-connected architecture
can improve the spectral efficiency by up to 90\%; replacing an active
D-RIS with an active BD-RIS having non-reciprocal fully-connected
architecture can improve the spectral efficiency by around 24\%. More
importantly, to achieve the same performance, the required number
of elements for active BD-RIS is much smaller than that for active
D-RIS. For example, we can observe from Fig. 7(a) that, to achieve
the spectral efficiency of around 17 b/s/Hz, the required number of
elements for active D-RIS is around 64, while that for reciprocal
active BD-RIS with group-connected architectures ($N_{\mathrm{G}}=2$)
reduces to 24. For another example, we observe from Fig. 7(b) that,
to achieve the spectral efficiency of 24 b/s/Hz, the required number
of elements for active D-RIS is 112 and that for reciprocal active
BD-RIS with group-connected architectures reduces to 48. These results
highlight the benefit of active BD-RIS in providing sufficient performance
gains within limited aperture.

\section{Conclusion and Discussions}

In this work, we investigate the active BD-RIS including the modeling,
architecture design, and optimization, to address the concern of multiplicative
path loss from the passive BD-RISs. Compared with the active D-RIS
characterized by a diagonal scattering matrix, the proposed active
BD-RIS is characterized by a beyond-diagonal scattering matrix, providing
more degrees of freedom to enhance wireless systems.

We first analyze the active BD-RIS aided wireless communication system
using multiport network theory with scattering parameters and derive
a physical and EM compliant active BD-RIS aided communication model
which is general to account for the active reconfigurable impedance
network of active BD-RIS, impedance mismatching, and mutual coupling.
We also design two new active BD-RIS architectures, namely fully-
and group-connected active BD-RISs, which are respectively characterized
by a complex symmetric matrix and a block diagonal matrix where each
block is complex symmetric.

Using the proposed model and architecture, we investigate the active
BD-RIS aided SISO system and derive the closed-form optimal solution
and scaling law of SNR, which shows that the maximum SNR gain of active
BD-RIS over active D-RIS in SISO system can be up to 1.62 and that
the active BD-RIS maintains its benefit over passive BD-RIS until
an extremely large number of elements. We also investigate the active
BD-RIS aided MIMO system and propose an iterative algorithms based
on QCQP to maximize the spectral efficiency. Numerical results show
that the active BD-RIS achieves higher spectral efficiency than the
active/passive D-RIS and passive BD-RIS. For example, to achieve the
same spectral efficiency, the number of elements required by active
BD-RIS is less than half of that required by active D-RIS, showing
the superior performance of active BD-RIS.

Inspired by the above promising results of active BD-RIS, there are multiple future research directions worth exploring, some of which are elaborated below.
\begin{enumerate}[1)]
    \item Develop computational efficient optimization algorithms. Note that the computational complexity of the proposed algorithm is as high as the order of the cube of the number of active BD-RIS elements and group size. To reduce the computational complexity, modern learning-based algorithms can be developed to achieve better performance-computational complexity tradeoff. 
    \item Extend active BD-RIS to support various architectures and modes. The architecture design, optimization, and performance evaluation in this work are restricted to active BD-RIS with reflective mode. Analogous to passive BD-RIS, it is also meaningful to extend current model to support other architectures, such as tree/forest and band/stem-connected, with better performance-circuit complexity tradeoff, and to support hybrid/multi-sector modes with enlarged and highly directional wireless coverage. 
    \item Estimate active BD-RIS aided wireless channels with low overhead and estimation error. Different from passive BD-RIS aided systems where cascaded channel knowledge can be sufficient for beamforming and data transmission, active BD-RIS aided systems require separate transmitter-RIS and RIS-user channel knowledge due to the additional dynamic noise. Although installing a few active devices can help to acquire separate channels, it remains unexplored how the additional dynamic noise will impact the estimation performance and how the amplification capability of active BD-RIS can help to reduce the channel estimation error. 
    \item Model and analyze hardware impairments of active BD-RIS. Practical active BD-RIS inevitably suffers impairments from various perspectives, such as insertion loss of the passive reconfigurable impedance network, amplification limitation and nonlinearity of amplifiers \cite{Gavriilidis2025}, and mutual coupling between antenna elements. Therefore, it is meaningful to accurately model hardware impairments of active BD-RIS and develop effective compensation strategies to guarantee the stability and show the effectiveness of practical active BD-RIS. 
\end{enumerate}

\bibliographystyle{IEEEtran}
\bibliography{RIS,REF}

\begin{thebibliography}{10}
\providecommand{\url}[1]{#1}
\csname url@samestyle\endcsname
\providecommand{\newblock}{\relax}
\providecommand{\bibinfo}[2]{#2}
\providecommand{\BIBentrySTDinterwordspacing}{\spaceskip=0pt\relax}
\providecommand{\BIBentryALTinterwordstretchfactor}{4}
\providecommand{\BIBentryALTinterwordspacing}{\spaceskip=\fontdimen2\font plus
\BIBentryALTinterwordstretchfactor\fontdimen3\font minus
  \fontdimen4\font\relax}
\providecommand{\BIBforeignlanguage}[2]{{%
\expandafter\ifx\csname l@#1\endcsname\relax
\typeout{** WARNING: IEEEtran.bst: No hyphenation pattern has been}%
\typeout{** loaded for the language `#1'. Using the pattern for}%
\typeout{** the default language instead.}%
\else
\language=\csname l@#1\endcsname
\fi
#2}}
\providecommand{\BIBdecl}{\relax}
\BIBdecl

\bibitem{Mu2024}
X.~Mu, J.~Xu, Y.~Liu, and L.~Hanzo, ``Reconfigurable intelligent surface-aided
  near-field communications for {6G}: Opportunities and challenges,''
  \emph{IEEE Veh. Technol. Mag.}, vol.~19, no.~1, pp. 65--74, 2024.

\bibitem{di2020smart}
M.~Di~Renzo, A.~Zappone, M.~Debbah, M.-S. Alouini, C.~Yuen, J.~De~Rosny, and
  S.~Tretyakov, ``Smart radio environments empowered by reconfigurable
  intelligent surfaces: How it works, state of research, and the road ahead,''
  \emph{IEEE J. Sel. Areas Commun.}, vol.~38, no.~11, pp. 2450--2525, 2020.

\bibitem{li2023reconfigurable}
H.~Li, S.~Shen, M.~Nerini, and B.~Clerckx, ``Reconfigurable intelligent
  surfaces 2.0: Beyond diagonal phase shift matrices,'' \emph{IEEE Commun.
  Mag.}, vol.~62, no.~3, pp. 102--108, 2024.

\bibitem{Li2025c}
H.~Li, M.~Nerini, S.~Shen, and B.~Clerckx, ``A tutorial on beyond-diagonal
  reconfigurable intelligent surfaces: Modeling, architectures, system design
  and optimization, and applications,'' \emph{IEEE Commun. Surveys \& Tuts.},
  pp. 1--1, 2025.

\bibitem{shen2021}
S.~Shen, B.~Clerckx, and R.~Murch, ``Modeling and architecture design of
  reconfigurable intelligent surfaces using scattering parameter network
  analysis,'' \emph{IEEE Trans. Wireless Commun.}, vol.~21, no.~2, pp.
  1229--1243, 2021.

\bibitem{Li2023a}
H.~Li, S.~Shen, and B.~Clerckx, ``A dynamic grouping strategy for beyond
  diagonal reconfigurable intelligent surfaces with hybrid transmitting and
  reflecting mode,'' \emph{IEEE Trans. Veh. Technol.}, vol.~72, no.~12, pp.
  16\,748--16\,753, 2023.

\bibitem{Nerini2024c}
M.~Nerini, S.~Shen, and B.~Clerckx, ``Static grouping strategy design for
  beyond diagonal reconfigurable intelligent surfaces,'' \emph{IEEE Commun.
  Lett.}, 2024.

\bibitem{Nerini2024b}
M.~Nerini, S.~Shen, H.~Li, and B.~Clerckx, ``Beyond diagonal reconfigurable
  intelligent surfaces utilizing graph theory: Modeling, architecture design,
  and optimization,'' \emph{IEEE Trans. Wireless Commun.}, 2024.

\bibitem{Wu2025a}
Z.~Wu and B.~Clerckx, ``Beyond diagonal {RIS} in multiuser {MIMO}: Graph
  theoretic modeling and optimal architectures with low complexity,''
  \emph{arXiv:2502.16509}, 2025.

\bibitem{li2022beyond}
H.~Li, S.~Shen, and B.~Clerckx, ``Beyond diagonal reconfigurable intelligent
  surfaces: From transmitting and reflecting modes to single-, group-, and
  fully-connected architectures,'' \emph{IEEE Trans. Wireless Commun.},
  vol.~22, no.~4, pp. 2311--2324, 2022.

\bibitem{Li2023}
------, ``Beyond diagonal reconfigurable intelligent surfaces: A multi-sector
  mode enabling highly directional full-space wireless coverage,'' \emph{IEEE
  J. Sel. Areas Commun.}, vol.~41, no.~8, pp. 2446--2460, 2023.

\bibitem{Azarbahram2025a}
A.~Azarbahram, O.~L. Lopez, B.~Clerckx, M.~Di~Renzo, and M.~Latva-Aho,
  ``Beamforming and waveform optimization for {RF} wireless power transfer with
  beyond diagonal reconfigurable intelligent surfaces,''
  \emph{arXiv:2502.19176}, 2025.

\bibitem{Wang2024e}
B.~Wang, H.~Li, S.~Shen, Z.~Cheng, and B.~Clerckx, ``A dual-function
  radar-communication system empowered by beyond diagonal reconfigurable
  intelligent surface,'' \emph{IEEE Trans. Commun.}, 2024.

\bibitem{Liu2024a}
Z.~Liu, Y.~Liu, S.~Shen, Q.~Wu, and Q.~Shi, ``Enhancing {ISAC} network
  throughput using beyond diagonal {RIS},'' \emph{IEEE Wireless Commun. Lett.},
  2024.

\bibitem{Guang2024}
Z.~Guang, Y.~Liu, Q.~Wu, W.~Wang, and Q.~Shi, ``Power minimization for {ISAC}
  system using beyond diagonal reconfigurable intelligent surface,'' \emph{IEEE
  Trans. Veh. Technol.}, 2024.

\bibitem{Soleymani2023}
M.~Soleymani, I.~Santamaria, E.~A. Jorswieck, and B.~Clerckx, ``Optimization of
  rate-splitting multiple access in beyond diagonal {RIS}-assisted {URLLC}
  systems,'' \emph{IEEE Trans. Wireless Commun.}, vol.~23, no.~5, pp.
  5063--5078, 2023.

\bibitem{Li2024c}
H.~Li, S.~Shen, and B.~Clerckx, ``Synergizing beyond diagonal reconfigurable
  intelligent surface and rate-splitting multiple access,'' \emph{IEEE Trans.
  Wireless Commun.}, vol.~23, no.~8, pp. 8717--8729, 2024.

\bibitem{DeSena2024}
A.~S. De~Sena, M.~Rasti, N.~H. Mahmood, and M.~Latva-Aho, ``Beyond diagonal
  {RIS} for multi-band multi-cell {MIMO} networks: A practical
  frequency-dependent model and performance analysis,'' \emph{IEEE Trans.
  Wireless Commun.}, 2024.

\bibitem{Hua2024}
T.~D. Hua, M.~Mohammadi, H.~Q. Ngo, and M.~Matthaiou, ``Cell-free massive {MIMO
  SWIPT} with beyond diagonal reconfigurable intelligent surfaces,'' in
  \emph{IEEE Wireless Commun. Network. Conf. (WCNC)}.\hskip 1em plus 0.5em
  minus 0.4em\relax IEEE, 2024, pp. 1--6.

\bibitem{Li2025b}
Y.~Li, J.~Zheng, B.~Xu, Y.~Zhu, J.~Zhang, and B.~Ai, ``Beamforming design for
  beyond diagonal {RIS}-aided cell-free massive {MIMO} systems,''
  \emph{arXiv:2503.07189}, 2025.

\bibitem{Li2025}
H.~Li, M.~Nerini, S.~Shen, and B.~Clerckx, ``Beyond diagonal reconfigurable
  intelligent surfaces in wideband {OFDM} communications: Circuit-based
  modeling and optimization,'' \emph{IEEE Trans. Wireless Commun.}, 2025.

\bibitem{Li2024a}
H.~Li, S.~Shen, Y.~Zhang, and B.~Clerckx, ``Channel estimation and beamforming
  for beyond diagonal reconfigurable intelligent surfaces,'' \emph{IEEE Trans.
  Signal Process.}, 2024.

\bibitem{Wang2025}
R.~Wang, S.~Zhang, B.~Clerckx, and L.~Liu, ``Low-overhead channel estimation
  framework for beyond diagonal reconfigurable intelligent surface assisted
  multi-user {MIMO} communication,'' \emph{arXiv:2504.10911}, 2025.

\bibitem{Ming2025a}
Z.~Ming, S.~Shen, J.~Rao, Z.~Li, J.~Zhang, C.-Y. Chiu, and R.~Murch, ``A hybrid
  transmitting and reflecting beyond-diagonal reconfigurable intelligent
  surface with independent beam control and power splitting,'' \emph{IEEE
  Trans. Microwave Theory Tech.}, vol.~73, no.~12, pp. 10\,865--10\,883, 2025.

\bibitem{Long2021}
R.~Long, Y.-C. Liang, Y.~Pei, and E.~G. Larsson, ``Active reconfigurable
  intelligent surface-aided wireless communications,'' \emph{IEEE Trans.
  Wireless Commun.}, vol.~20, no.~8, pp. 4962--4975, 2021.

\bibitem{Khoshafa2021}
M.~H. Khoshafa, T.~M.~N. Ngatched, M.~H. Ahmed, and A.~R. Ndjiongue, ``Active
  reconfigurable intelligent surfaces-aided wireless communication system,''
  \emph{IEEE Commun. Lett.}, vol.~25, no.~11, pp. 3699--3703, 2021.

\bibitem{Ahmed2025}
M.~Ahmed, S.~Raza, A.~Amin~Soofi, F.~Khan, W.~Ullah~Khan, S.~Zain Ul~Abideen,
  F.~Xu, and Z.~Han, ``Active reconfigurable intelligent surfaces: Expanding
  the frontiers of wireless communication-a survey,'' \emph{IEEE Commun.
  Surveys \& Tuts.}, vol.~27, no.~2, pp. 839--869, 2025.

\bibitem{Zhang2023}
Z.~Zhang, L.~Dai, X.~Chen, C.~Liu, F.~Yang, R.~Schober, and H.~V. Poor,
  ``Active {RIS} vs. passive {RIS}: Which will prevail in {6G}?'' \emph{IEEE
  Trans. Commun.}, vol.~71, no.~3, pp. 1707--1725, 2023.

\bibitem{Liu2022b}
K.~Liu, Z.~Zhang, L.~Dai, S.~Xu, and F.~Yang, ``Active reconfigurable
  intelligent surface: Fully-connected or sub-connected?'' \emph{IEEE Commun.
  Lett.}, vol.~26, no.~1, pp. 167--171, 2022.

\bibitem{Liu2022c}
Y.~Liu, Y.~Ma, M.~Li, Q.~Wu, and Q.~Shi, ``Spectral efficiency maximization for
  double-faced active reconfigurable intelligent surface,'' \emph{IEEE Trans.
  Signal Process.}, vol.~70, pp. 5397--5412, 2022.

\bibitem{Ma2023}
Y.~Ma, M.~Li, Y.~Liu, Q.~Wu, and Q.~Liu, ``Active reconfigurable intelligent
  surface for energy efficiency in {MU}-{MISO} systems,'' \emph{IEEE Trans.
  Veh. Technol.}, vol.~72, no.~3, pp. 4103--4107, 2023.

\bibitem{Zargari2022}
S.~Zargari, A.~Hakimi, C.~Tellambura, and S.~Herath, ``Multiuser {MISO}
  {PS-SWIPT} systems: Active or passive {RIS}?'' \emph{IEEE Wireless Commun.
  Lett.}, vol.~11, no.~9, pp. 1920--1924, 2022.

\bibitem{Ren2023}
H.~Ren, Z.~Chen, G.~Hu, Z.~Peng, C.~Pan, and J.~Wang, ``Transmission design for
  active {RIS}-aided simultaneous wireless information and power transfer,''
  \emph{IEEE Wireless Commun. Lett.}, vol.~12, no.~4, pp. 600--604, 2023.

\bibitem{Chen2024a}
P.~Chen, L.~Yang, J.~Wang, W.~Xie, X.~Li, Z.~Yan, and H.~Liu, ``Covert
  transmission and physical-layer security of active {RIS}-{RS}-{NOMA}-aided
  communication systems,'' \emph{IEEE Internet Things J.}, vol.~11, no.~19, pp.
  31\,507--31\,520, 2024.

\bibitem{Gong2024}
C.~Gong, H.~Li, S.~Hao, K.~Long, and X.~Dai, ``Active {RIS} enabled secure
  {NOMA} communications with discrete phase shifting,'' \emph{IEEE Trans.
  Wireless Commun.}, vol.~23, no.~4, pp. 3493--3506, Apr. 2024.

\bibitem{Peng2022}
Z.~Peng, R.~Weng, Z.~Zhang, C.~Pan, and J.~Wang, ``Active reconfigurable
  intelligent surface for mobile edge computing,'' \emph{IEEE Wireless Commun.
  Lett.}, vol.~11, no.~12, pp. 2482--2486, 2022.

\bibitem{Li2024d}
Y.~Li, Y.~Zou, H.~Hui, J.~Zhu, and B.~Ning, ``Improving computing capability
  for active {RIS}-assisted {NOMA}-{MEC} networks,'' \emph{IEEE Wireless
  Commun. Lett.}, vol.~13, no.~4, pp. 939--943, 2024.

\bibitem{Niu2023}
H.~Niu, Z.~Lin, K.~An, J.~Wang, G.~Zheng, N.~Al-Dhahir, and K.-K. Wong,
  ``Active {RIS} assisted rate-splitting multiple access network: Spectral and
  energy efficiency tradeoff,'' \emph{IEEE J. Sel. Areas Commun.}, vol.~41,
  no.~5, pp. 1452--1467, 2023.

\bibitem{Chian2024}
D.-M. Chian, F.-J. Chen, Y.-C. Chang, C.-K. Wen, C.-H. Wu, F.-K. Wang, K.-K.
  Wong, and C.-B. Chae, ``Active {RIS}-assisted {MIMO}-{OFDM} system: Analyses
  and prototype measurements,'' \emph{IEEE Commun. Lett.}, vol.~28, no.~1, pp.
  208--212, 2024.

\bibitem{Zhang2023a}
J.~Zhang, Z.~Li, and Z.~Zhang, ``Wideband active {RIS}s: Architecture,
  modeling, and beamforming design,'' \emph{IEEE Commun. Lett.}, vol.~27,
  no.~7, pp. 1899--1903, 2023.

\bibitem{Salem2023}
A.~A. Salem, M.~H. Ismail, and A.~S. Ibrahim, ``Active reconfigurable
  intelligent surface-assisted {MISO} integrated sensing and communication
  systems for secure operation,'' \emph{IEEE Trans. Veh. Technol.}, vol.~72,
  no.~4, pp. 4919--4931, 2023.

\bibitem{Sun2024a}
G.~Sun, H.~Shi, B.~Shang, and W.~Hao, ``Secure transmission for active
  {RIS}-assisted {THz} {ISAC} systems with delay alignment modulation,''
  \emph{IEEE Commun. Lett.}, vol.~28, no.~5, pp. 1019--1023, 2024.

\bibitem{Peng2024a}
Z.~Peng, R.~Liu, C.~Pan, Z.~Zhang, and J.~Wang, ``Energy minimization for
  active {RIS}-aided {UAV}-enabled {SWIPT} systems,'' \emph{IEEE Commun.
  Lett.}, vol.~28, no.~6, pp. 1372--1376, 2024.

\bibitem{Gavriilidis2025}
P.~Gavriilidis, D.~Mishra, B.~Smida, E.~Basar, C.~Yuen, and G.~C.
  Alexandropoulos, ``Active reconfigurable intelligent surfaces: Circuit
  modeling and reflection amplification optimization,'' \emph{IEEE Open J.
  Commun. Soc.}, vol.~6, pp. 5693--5711, 2025.

\bibitem{yumeng2023}
Y.~Zhang and B.~Clerckx, ``Waveform design for wireless power transfer with
  power amplifier and energy harvester non-linearities,'' \emph{IEEE
  Transactions on Signal Processing}, vol.~71, pp. 2638--2653, 2023.

\bibitem{francesco2015}
F.~Amato, C.~W. Peterson, B.~P. Degnan, and G.~D. Durgin, ``A 45 {$\mu$W} bias
  power, 34 {dB} gain reflection amplifier exploiting the tunneling effect for
  {RFID} applications,'' in \emph{IEEE Int. Conf. RFID (RFID)}, 2015, pp.
  137--144.

\bibitem{jongwon2017}
J.~Lee and K.~Yang, ``{RF} power analysis on 5.8 {GHz} low-power amplifier
  using resonant tunneling diodes,'' \emph{IEEE Microwave Wireless Components
  Lett.}, vol.~27, no.~1, pp. 61--63, 2017.

\bibitem{Seiran2017}
S.~Khaledian, F.~Farzami, D.~Erricolo, and B.~Smida, ``A full-duplex
  bidirectional amplifier with low {DC} power consumption using tunnel
  diodes,'' \emph{IEEE Microwave Wireless Components Lett.}, vol.~27, no.~12,
  pp. 1125--1127, 2017.

\bibitem{rao2023}
J.~Rao, Y.~Zhang, S.~Tang, Z.~Li, C.-Y. Chiu, and R.~Murch, ``An active
  reconfigurable intelligent surface utilizing phase-reconfigurable reflection
  amplifiers,'' \emph{IEEE Trans. Microwave Theory Techn.}, vol.~71, no.~7, pp.
  3189--3202, 2023.

\bibitem{horn2012matrix}
R.~A. Horn and C.~R. Johnson, \emph{Matrix analysis}.\hskip 1em plus 0.5em
  minus 0.4em\relax Cambridge university press, 2012.

\bibitem{ivrlavc2010toward}
M.~T. Ivrla{\v{c}} and J.~A. Nossek, ``Toward a circuit theory of
  communication,'' \emph{IEEE Trans. Circuits Syst. I, Reg. Papers}, vol.~57,
  no.~7, pp. 1663--1683, 2010.

\bibitem{nerini2023closed}
M.~Nerini, S.~Shen, and B.~Clerckx, ``Closed-form global optimization of beyond
  diagonal reconfigurable intelligent surfaces,'' \emph{IEEE Trans. Wireless
  Commun.}, vol.~23, no.~2, pp. 1037--1051, 2023.

\bibitem{santamaria2023snr}
I.~Santamaria, M.~Soleymani, E.~Jorswieck, and J.~Guti{\'e}rrez, ``{SNR}
  maximization in beyond diagonal {RIS}-assisted single and multiple antenna
  links,'' \emph{IEEE Signal Process. Lett.}, vol.~30, pp. 923--926, 2023.

\bibitem{wu2021intelligent}
Q.~Wu, S.~Zhang, B.~Zheng, C.~You, and R.~Zhang, ``Intelligent reflecting
  surface-aided wireless communications: A tutorial,'' \emph{IEEE Trans.
  Commun.}, vol.~69, no.~5, pp. 3313--3351, 2021.

\bibitem{christensen2008weighted}
S.~S. Christensen, R.~Agarwal, E.~De~Carvalho, and J.~M. Cioffi, ``Weighted
  sum-rate maximization using weighted {MMSE} for {MIMO-BC} beamforming
  design,'' \emph{IEEE Trans. Wireless Commun.}, vol.~7, no.~12, pp.
  4792--4799, 2008.

\bibitem{santamaria2024mimo}
I.~Santamaria, M.~Soleymani, E.~Jorswieck, and J.~Guti{\'e}rrez, ``{MIMO}
  capacity maximization with beyond-diagonal {RIS},'' in \emph{IEEE 25th Int.
  Workshop Signal Process. Advances Wireless Commun. (SPAWC)}.\hskip 1em plus
  0.5em minus 0.4em\relax IEEE, 2024, pp. 936--940.

\end{thebibliography}

\end{document}